\documentclass[3p, times, reprint, 10pt]{elsarticle}

\usepackage{amsmath,amssymb,amsfonts}
\usepackage{algorithmic}
\usepackage{graphicx}
\usepackage{epstopdf}
\usepackage{textcomp}
\usepackage{multirow}
\usepackage{booktabs}
\usepackage{amsmath}
\usepackage{lineno,hyperref}

\modulolinenumbers[1]

\journal{xxxx}

\begin{document}

\begin{frontmatter}

\title{Towards a Unified Approach to Electromagnetic Analysis of Objects Embedded in Multilayers}

\author[rvt]{Xiaochao Zhou}

\author[rvt]{Zekun Zhu}
\author[rvt]{Shunchuan Yang \corref{cor1}}
\cortext[cor1]{Corresponding author}
\ead{scyang@buaa.edu.cn}

\address[rvt]{Beihang University, Xueyuan Road 37, Haidian District, Beijing, China}

\begin{abstract}
According to the surface equivalence theorem {\cite{CHUSTRATT}}, any enclosed surface with an electric and magnetic currents enforced on it can generate exactly the same fields as the original problem. The problem is how to construct the current sources. In this paper, an efficient and accurate unified approach with an equivalent current density enforced on the outermost boundary is proposed to solve transverse magnetic (TM) scattering problems by objects embedded in multilayers. In the proposed approach, an equivalent current density is derived after the surface equivalence theorem is recursively applied on each boundary from innermost to outermost interfaces.  Then, the objects are replaced by the background medium and the equivalent electric current density only on the outermost boundary is derived. The scattering problems by objects embedded in multilayers can be solved with the electric field integral equation (EFIE). Compared with the Poggio-Miller-Chan-Harrington-Wu-Tsai (PMCHWT) and other dual source formulations, the proposed approach shows significant benefits:  only the surface electric current density instead of both the electric and magnetic current densities is required to model the complex objects in the equivalent  problem. Furthermore, the electric current density is only enforced on the outermost boundary of objects. Therefore, the overall count of unknowns can be significantly reduced. At last, several numerical experiments are performed to validate its accuracy and efficiency. Although overhead is required to construct the intermediate matrixes, the overall performance improvement is still significant  as numerical results are shown. Therefore, it shows great potential useful in the practical engineering applications.
\end{abstract}


\end{frontmatter}


\section{Introduction}
The objects embedded in multilayers are widely used in the practical engineering applications, such as multilayer integrated circuits \cite{INTERNETS}, thin layer coated fibers \cite{FIBERS}, power cables \cite{POWERCABLES}, coating aircrafts \cite{COATEDAIRCRAFTS}, and so on. There are various numerical methods proposed to model such objects. One of the powerful numerical tools is the method of moment (MOM) \cite{MOMBOOK} based on the surface integral equations (SIEs), which is widely used due to its unknowns residing on the interfaces of different homogenous media. Therefore, the overall count of unknowns is significantly less than those of partial differential equation (PDE) based methods, like the finite element method (FEM) \cite{FEM} and the finite-difference time-domain (FDTD) method \cite{FDTD}, which requires volumetric meshes.

There are various SIEs, like the Poggio-Miller-Chan-Harrington-Wu-Tsai (PMCHWT) formulation \cite{PMCHWT}, the combined tangential formulation (CTF) \cite{CTF}, can model objects embedded in multilayers. In these formulations, both the surface equivalent electric and magnetic current densities are introduced and several two-region problems are required to be solved simultaneously. Therefore, as the layer number increases, the overall count of unknowns will greatly increase.

To mitigate the problem above, a number of efforts are made to introduce single-source formulations. In \cite{SVS}, a single-source surface-volume-surface formulation is proposed to model penetrable objects. In this formulation, the volumetric integral operator is mapped from surface to volume operator and then back from volume to surface operator. Therefore, significant efficiency improvement can be obtained. However, compared with the SIEs, it still requires to evaluate the volume integral operator. Various other single-source formulations based on the surface equivalence theorem are proposed in \cite{SINGLESORC1, SINGLESORC2, SINGLESORC3, SINGLESORC4}. Only single current source is required in those formulations. Therefore, more efficient formulations can be obtained compared with their dual source counterparts. Through combing the magnetic field integral equation (MFIE) inside the conductor and electric field integral field (EFIE) outside the conductor and mathematically eliminating the magnetic current, a generalized impedance boundary condition (GIBC) is proposed to model the interconnects \cite{GIBC}. In \cite{DSAO}, a single-source formulation based on the differential surface admittance operator (DSAO) to model high-speed interconnects is proposed. The DSAO is derived through the surface equivalent theorem and electric fields in the equivalent problem are enforced to equal to those in the original problem. Its capability is further enhanced to model arbitrarily shaped interconnects \cite{DSAOARBSHAPEDCIM}, circular solid and hollow cables \cite{DSAOARBSHAPEHOLLOW}, circular hole cables \cite{DSAOHOLLOW}, three dimensional scattering \cite{DSAOARBSHAPEDO}.  There are several  equivalence principle algorithms (EPAs) proposed in \cite{EPA}, \cite{EPAHOF}, \cite{EPAMPA}, in which a fictitious  surface is constructed and {\it{both}} the equivalent surface electric and magnetic currents are enforced on the fictitious boundary. Those show better conditioning and possible less number of knowns compared with direct solution of the MOM. Especially, in \cite{DSAOARBSHAPEARRAY}, each element in a large antenna array is replaced by a fictitious enclosed surface and a {\it{single}} equivalent surface current density is introduced to ensure the fields unchanged, which is similar to the EPA \cite{EPA}. This formulation shows better conditioning in the coefficient matrix, therefore, better convergence properties and great performance improvement compared with the direct MOM. However, in those formulations, they are only applicable for penetrable objects and those two-regions problems are required to be solved simultaneously as the PMCHWT formulation. They still suffer from the problem stated above.

In this paper, we proposed a unified single-source SIE incorporated the DSAO to solve two dimensional transverse magnetic (TM) scattering problems by objects  embedded in multilayers to address the problem above. The equivalent theorem is recursively applied  from the innermost to outermost boundaries. An equivalent object is then derived in the proposed formulation, which is filled with the background medium, and {\it an electric current density} enforced on the original {\it outermost} boundary. It should be noted that the proposed approach is significantly different from the approach in \cite{RECURSIVE}, which is based on the definition of recursive Green function. In \cite{RECURSIVE}, the recursive Green function is defined through recursively  combing unit fictitious cells and then is used to solve the entire problem. However, the proposed approach directly uses the surface equivalent theorem on each interface and derives the equivalent current density. Compared with the PMCHWT formulation \cite{PMCHWT}, other two-region formulations \cite{CTF}, and the EPA based techniques \cite{EPA, EPAHOF, EPAMPA}, the overall count of unknowns of the proposed approach can be significantly reduced.  The formulation shows three features: (1) Only the single-source electric current is required. (2) The surface equivalent electric current density is only enforced on the outermost boundary rather than all the boundaries between different homogenous media. (3) If the multilayered cylindrical objects are involved, the proposed method can avoid the troublesome multilayer Green function evaluation, the proposed approach only requires to evaluate the free space Green function. Compared with those single-source formulations \cite{SINGLESORC1, SINGLESORC2, SINGLESORC3, SINGLESORC4} and GIBC \cite{GIBC}, the proposed approach can model objects embedded in multilayers and only apply the single current source on the outermost boundary. The approach proposed in this paper can significantly simplify the analysis and provide a unified approach to analyze objects embedded in multilayers and reduce the overall number of unknowns for complex objects. We reported some preliminary  results upon it in \cite{XiaoChao} and this paper extensively and rigorously analyzed the proposed approach.

The main contributions in this paper are summarized into three aspects as follows. (1) A novel and general SIE is proposed to model objects embedded in multilayers. In this formulation, the surface equivalence theorem is recursively applied from the innermost to outermost interfaces. Only surface equivalent sources are enforced on the outermost boundary. Compared with other SIE based techniques, which solve two-region problems simultaneously, each two region problem is solved separately. Therefore, several small problems rather than a large one including all the unknowns residing on all the interfaces are solved in the proposed formulation. In addition, the proposed formulation is better conditioned than the PMCHWT formulation since unknowns only reside on the outermost boundary and any possible geometry fine details are only implicitly included in the final system. Therefore, the proposed approach can improve the conditioning and thus the convergence property of the final linear system.  (2) The proposed approach is incorporated with the DSAO to further enhance the performance. Only an electric current density is required to be enforced on the outermost boundary of objects, which retain the fields the same as those in the original problem. Unknowns involved only the electric current density in the proposed SIE only reside on the outermost boundary.  Therefore, significant efficiency improvement can be obtained. (3) Two special  scenarios of the proposed approach are derived. One is that perfectly electric conductor (PEC) scatters embedded in multilayers. The other is that a fictitious boundary resides in the same medium.  These are commonly used in the practical engineering problems, like the antenna array modeling \cite{DSAOARBSHAPEARRAY} and circuit modeling \cite{GIBC}.

This paper is organized as follows. In Section 2,  the configuration of objects embedded in multilayers is described in detail. We proposed a new unified approach to model these objects in Section 3. We introduced a single source formulation incorporated with the DSAO for a single penetrable object in the free space. Then, we proposed a recursive approach for two layered embedded objects and further extended the proposed approach to model objects embedded in multilayers. Two special scenarios including PEC objects embedded in multilayers and extension in the same medium are derived based on the proposed approach. In Section 4, some remarks upon the performance  upon the proposed approach are made. Then, in Section 5,  its performance in terms of convergence property, accuracy and efficiency are numerically validated through three numerical examples. At last, we draw some conclusions in Section 6.

\section{Methodology}
\subsection{The Problem Configuration}
\begin{figure}
	\begin{minipage}[h]{0.5\linewidth}\label{FIG2A}
		\centerline{\includegraphics[scale=0.5]{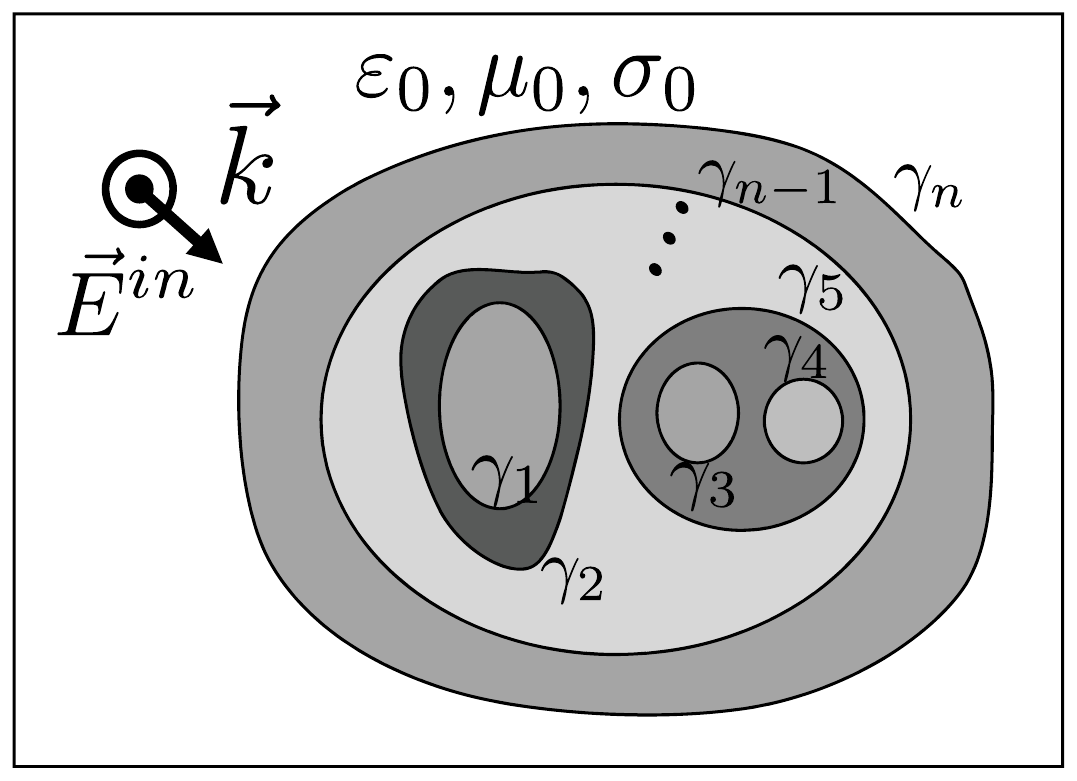}}
		\centerline{(a)}
	\end{minipage}
	\hfill
	\begin{minipage}[h]{0.5\linewidth}\label{FIG2B}
		\centerline{\includegraphics[scale=0.5]{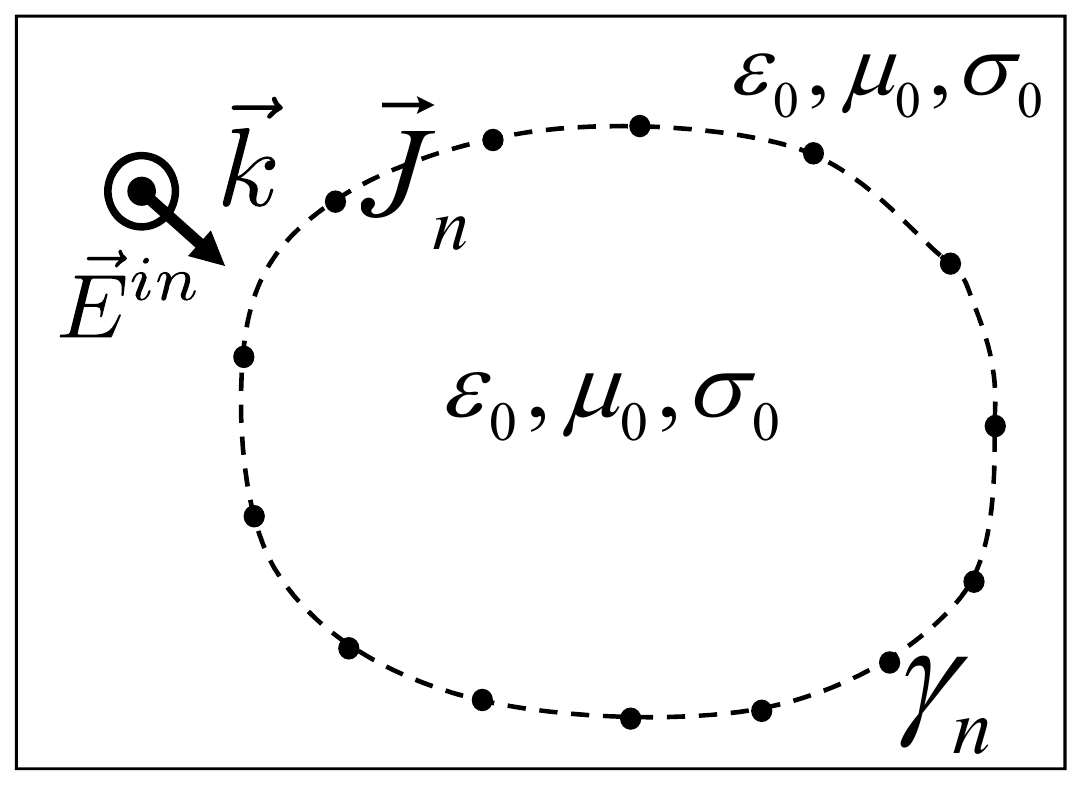}}
		\centerline{(b)}
	\end{minipage}
	\caption{(a) The original model and (b) the equivalent model with inner media replaced by its background medium and enforcing the surface electric current density $\vec J_n$ on the outermost boundary}
	\label{fig_2}
\end{figure}

In this paper, a two-dimensional TM scattering problem by objects embedded in multilayers is considered as shown in Fig. 1(a). To derive a unified approach to solve it, we resort to the equivalence theorem to obtain an equivalent model,  in which objects are replaced by the background medium and the exterior fields are exactly the same as those of the original model. Our goal is to introduce only one single surface equivalent electric or magnetic current density on the outermost boundary $\gamma_n$ as shown in Fig. 1(b) in the equivalent model to keep the fields in outermost region unchanged. We keep this goal in mind and will derive an equivalent current density enforced on the outmost boundary $\gamma_n$ from inner to exterior  layer by layer.  The original objects embedded in multilayers are shown in Fig. 1(a), where ${\varepsilon _i}$, ${\mu _i}$ denote the permittivity and the permeability of the $i$th layer medium, which is bounded by two adjacent boundaries, ${\gamma_{i-1}}$ and ${\gamma_i}$, respectively.

Before we derive the detailed formulations, let's make more remarks on the single source formulations. Our goal is to derive an equivalent model as shown in the Fig. 1(b), in which the equivalent electric current density ${{{\bf{\widehat J}}}_{i}}$ is introduced on the fictitious boundary, ${\gamma_i}$. According to the equivalent theorem \cite{EQUIVALENT}, the equivalent electric and magnetic current densities can be expressed as
\begin{align}{\label{EQUIJ}}
{{{\bf{\widehat J}}}_{i}}(\vec r) = {{\bf{H}}_{t_i}}(\vec r) - {{{\bf{\widehat H}}}_{t_i}}(\vec r),
\end{align}
\begin{align}{\label{EQUIM}}
{{{\bf{\widehat M}}}_{i}}(\vec r) = {{\bf{E}}_{t_i}}(\vec r) - {{{\bf{\widehat E}}}_{t_i}}(\vec r),
\end{align}
where $\vec r \in {\gamma_i}$, ${\bf{E}}_{t_i}(\vec r)$, $\widehat {\bf{E}}_{t_i}(\vec r)$, ${\bf{H}}_{t_i}(\vec r)$, $\widehat {\bf{H}}_{t_i}(\vec r)$ are the surface tangential electric and magnetic fields in the original and equivalent model, respectively. All quantities with $\quad\widehat{}\quad$ denote their values in the equivalent model.

Since the electric and magnetic fields in the equivalent model can be arbitrary, we can obtain the single electric current source by enforcing that ${\bf{E}}_{t_i}(\vec r) = \widehat {\bf{E}}_{t_i}(\vec r)$ and (\ref{EQUIJ}) and (\ref{EQUIM}) are rewritten as
\begin{equation}{\label{SINGLEJ}}
\widehat {\bf{J}}_{i}(\vec r) \neq {\bf{0}},  \widehat {\bf{M}}_{i}(\vec r) = {\bf{0}}.
\end{equation}
On the other hand, we can have the single magnetic current source by enforcing that ${\bf{H}}_{t_i}(\vec r) = \widehat {\bf{H}}_{t_i}(\vec r) $ and (\ref{EQUIJ}) and (\ref{EQUIM}) are expressed as
\begin{equation}{\label{SINGLEM}}
\widehat {\bf{M}}_{i}(\vec r) \neq {\bf{0}},\widehat {\bf{J}}_{i}(\vec r) = {\bf{0}}.
\end{equation}

Both (\ref{SINGLEJ}) and (\ref{SINGLEM}) can give us a single-source formulation. In this paper, we use (\ref{SINGLEJ}) to obtain the single source integral formulation and derive the surface equivalent electric current density using the contour integral method for objects embedded in multilayers. In the proposed approach, we will derive the DSAO  ${{\bf{Y}}_{s_n}}$ according to the surface equivalence theorem, which relates the outermost electric fields and currents of objects on $\gamma_n$.

\subsection{The Single-source  SIE for A Penetrable Object}
\begin{figure}
	\begin{minipage}[h]{0.5\linewidth}\label{FIG2A}
		\centerline{\includegraphics[scale=0.45]{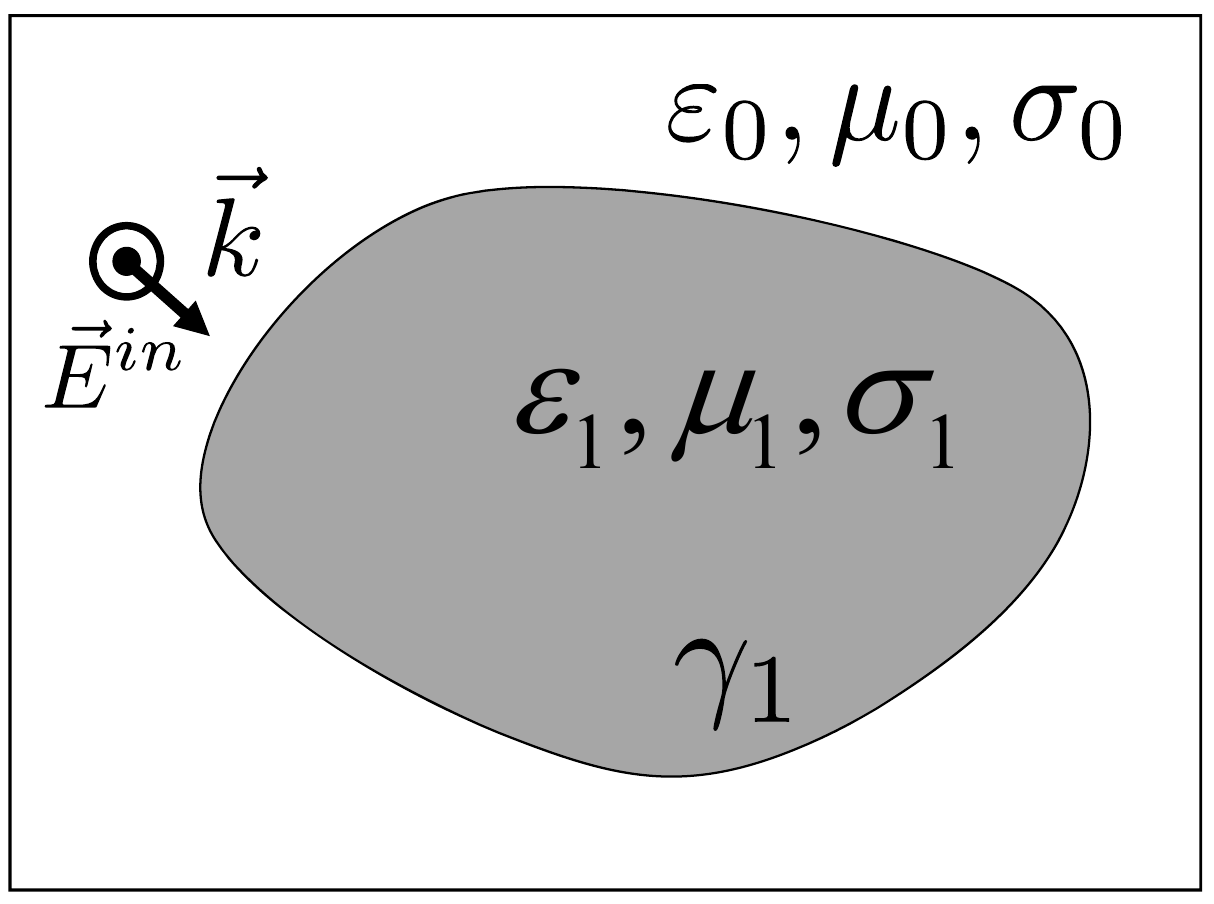}}
		\centerline{(a)}
	\end{minipage}
	\hfill
	\begin{minipage}[h]{0.5\linewidth}\label{FIG2B}
		\centerline{\includegraphics[scale=0.45]{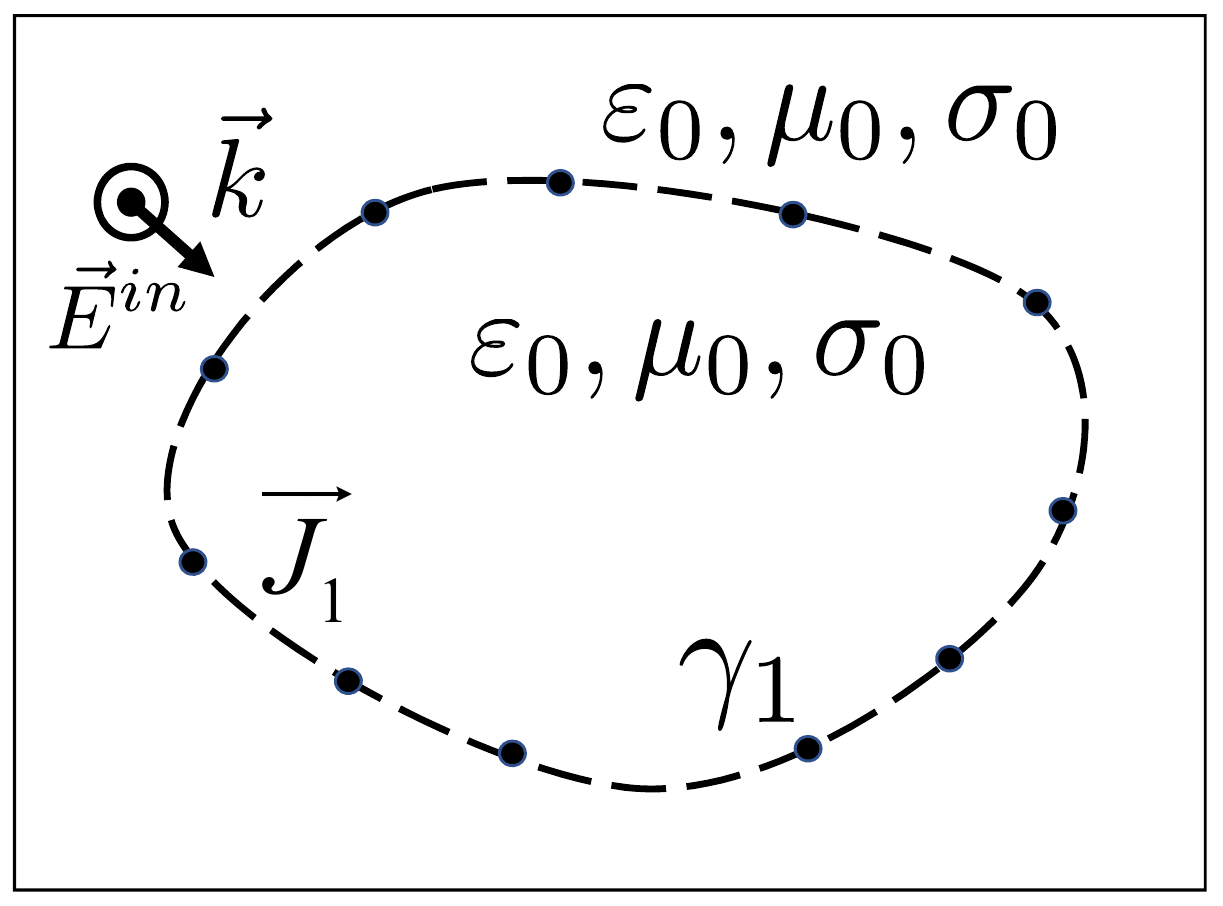}}
		\centerline{(b)}
	\end{minipage}
	\caption{(a) The original model for a single penetrable object and (b) the equivalent model}
	\label{fig_2}
\end{figure}

Let us first consider a single penetrable object with the permittivity ${\varepsilon _1}$, the permeability ${\mu _1}$ and the conductivity ${\sigma_1}$, respectively, as shown in Fig. 2(a). Its boundary is denoted as ${\gamma_1}$. The permittivity, the permeability and the conductivity of the background medium are ${\varepsilon _0}$, ${\mu _0}$ and ${\sigma_0}$, respectively. According to the equivalence theorem \cite{EQUIVALENT}, an equivalent model, in which the object is replaced by its surrounding medium and a surface equivalent electric current density is introduced on $\gamma_1$ as shown in the Fig. 2(b), can be obtained. Interested readers are referred to \cite{DSAOARBSHAPEDCIM} for more details.

Let's consider that the penetrable object does not include any sources, electric fields must satisfy the following scalar Helmholtz equation inside $\gamma_1$
\begin{equation}{\label{HELMHOLTZ1}}
{\nabla ^2}E_1 + k_1^2E_1 = 0,
\end{equation}
subject to the boundary condition
\begin{equation}{\label{BOUNDARYC1}}
{\left. {E_1(\vec r)} \right|_{\vec r \in {\gamma_1}}} = {\left. {\widehat E_1(\vec r)} \right|_{\vec r \in {\gamma_1}}},
\end{equation}
where $E_1$ and ${\widehat E_1}$ denote the electric field inside $\gamma_1$ for the original and equivalent models, respectively, and ${\left. {E_1(\vec r)} \right|_{\vec r \in {\gamma_1}}}$ and ${\left. {\widehat E_1(\vec r)} \right|_{\vec r \in {\gamma_1}}}$ denote their values on the inner side of $\gamma_1$.

({\ref{HELMHOLTZ1}}) can be solved through the second scalar Green theorem \cite{CIPBOOK}. Then, the electric field $E_1$  inside $\gamma_1$ can be expressed in terms of $E_1$ and its normal derivative on $\gamma_1$ as
\begin{equation}{\label{INTEGRALEQU1}}
{T}E_{1}(\vec r) = \oint_{{\gamma_1 }} {\left[ G_1(\vec r,\vec r')\frac{{\partial E_{1}(\vec r')}}{{\partial n'}}- {\frac{{\partial G_1(\vec r,\vec r')}}{{\partial n'}}E_{1}(\vec r')} \right]} dr',
\end{equation}
where the constant $T$ = 1/2 when the source and observation points are located on the same boundary, otherwise, $T =1$ and $G_1(\vec r,\vec r')$ is the Green function expressed as $G_1(\vec r,\vec r') = -j H_0^{(2)}({k_1}{\rho})/4$, where $j = \sqrt{-1}$, $k_1$ is the wavenumber in the penetrable object and $H_0^{(2)}(\cdot)$ is the {\it zero}th-order Hankel function of the second kind. In addition, the tangential magnetic field relates to the electric field on $\gamma_1$ through the Poincare-Steklov operator \cite{DSAO}
\begin{equation}{\label{HTM1}}
{H_{1}}(\vec r) = \frac{1}{{j{\omega}{\mu _1}}}{\left. { {\frac{{\partial {E_{1}}(\vec r)}}{{\partial n}}} } \right|_{\vec r \in {\gamma_1}}},
\end{equation}
where ${{\mu _1}}$ is the permeability of the object. We discretize $\gamma_1$ into $m_1$ segments and use the pulse basis function to expand $E_{1}$ and $H_{1}$ on $\gamma_1$ in (\ref{INTEGRALEQU1}) and (\ref{HTM1}) as
\begin{equation}{\label{BOUNDARYE1}}
{E_1}(\vec r) = \sum\limits_{n = 1}^{{m_1}} {{e_n}{f _n}(\vec r)},
\end{equation}
\begin{equation}{\label{BOUNDARYH1}}
{H_1}(\vec r) = \sum\limits_{n = 1}^{{m_1}} {{h_n}{f _n}(\vec r)},
\end{equation}
where ${{f _n}(\vec r)}$ denotes the $n$th basis function. We use the Galerkin scheme to test (\ref{INTEGRALEQU1}) and (\ref{HTM1}) at each segment of $\gamma_1$ and collect all $E_{1}$ and $H_{1}$ expansion coefficients into two column vectors $\bf{E_1}$ and $\bf{H_1}$ as
\begin{equation}{\label{BOUNDARYC}}
{{\bf{E}}_1} = {\left[ {\begin{array}{*{20}{c}}
		{e_{11}}&{e_{12}}&{...}&{{e_{1m_1}}}
		\end{array}} \right]^T},
\end{equation}
\begin{equation}{\label{BOUNDARYC}}
{{\bf{H}}_1} = {\left[ {\begin{array}{*{20}{c}}
		{h_{11}}&{h_{12}}&{...}&{{h_{1m_1}}}
		\end{array}} \right]^T}.
\end{equation}

Then, (\ref{INTEGRALEQU1}) can be rewritten into the matrix form as
\begin{equation}{\label{PU}}
{\frac{1}{2}{\bf{L}}_1}{{\bf{E}}_1} = {\bf{P}}_1^{(1)}{{\bf{H}}_1} + {\bf{U}}_1^{(1)}{{\bf{E}}_1},
\end{equation}
where $\mathbf{L}_1$ is a diagonal matrix with the length of each segment as its entities on $\gamma_1$, the subscript $1$ denotes that the testing procedure is applied on $\gamma_1$, and the superscript $(1)$ denotes the equivalence theorem applied on $\gamma_1$ in the next subsection, and elements of ${\bf{P}}_1^{(1)}$ and ${\bf{U}}_1^{(1)}$ are expressed as
\begin{equation}{\label{U1}}
{\left[ {{\bf{U}}_1^{(1)}} \right]_{m,n}} = \int_{{\gamma _{1_m}}} {\int_{{\gamma _{1_n}}} {{k_1}\frac{{{{\vec d}_m} \cdot \hat n'}}{{{d_m}}}} G'_1(\vec r,\vec r')dr'} dr,
\end{equation}
\begin{equation}{\label{P1}}
{\left[ {{\bf{P}}_1^{(1)}} \right]_{m,n}} = \int_{{\gamma _{1_m}}} {\int_{{\gamma _{1_n}}} {{\omega }{\mu _1}j{G_1}(\vec r,\vec r')dr'}}dr.
\end{equation}
where $G_{1}^{'}=-jH_{1}^{\left( 2 \right)}\left( k_1\rho \right) /4$.
Then, through inversing the square matrix ${{\bf{P}}_1^{(1)}}$, we obtain the surface admittance operator (SAO) ${\bf{Y}}_1$ \cite{DSAOARBSHAPEDCIM} as
\begin{equation}{\label{Y1}}
{{\bf{H}}_1} = \underbrace {{{[{\bf{P}}_1^{(1)}]}^{ - 1}}({\frac{1}{2}{\bf{L}}_1} - {\bf{U}}_1^{(1)})}_{{{\bf{Y}}_1}}{{\bf{E}}_1}.
\end{equation}

When all the parameters are replaced by those of its surrounding medium, the equivalent model is obtained as shown in Fig. 2(b). The ${{\widehat H}_1}(\vec r)$ is expanded through the pulse basis function as
\begin{equation}{\label{BOUNDARYH11}}
{{\widehat H}_1}(\vec r) = \sum\limits_{n = 1}^{{m_1}} {{{\widehat h}_n}{f _n}(\vec r)}.
\end{equation}

With the similar procedure in the original model, for the equivalent model we can obtain
\begin{equation}{\label{EQUY1}}
{\widehat {\bf{H}}_1} = \underbrace {{[{{\widehat {\bf{P}}}_1}^{(1)}]}^{ - 1}({\frac{1}{2}{\bf{L}}_1} - \widehat {\bf{U}}_1^{(1)})}_{{{\widehat {\bf{Y}}}_1}}{{\bf{E}}_1},
\end{equation}
where
\begin{equation}{\label{BOUNDARYC}}
{{{\bf{\widehat H}}}_1} = {\left[ {\begin{array}{*{20}{c}}
		{{{\widehat h}_{11}}}&{\widehat h_{12}}&{...}&{{{\widehat h}_{1m_1}}}
		\end{array}} \right]^T}
\end{equation}
and ${\widehat{\bf{H}}}_1$ denotes the discretized magnetic field expansion coefficient vector in the equivalent model.

The surface equivalent current density on $\gamma_1$ is expanded through pulse functions, and the coefficients are extracted into the column vector ${{\bf{J}}_1}$,
\begin{equation}{\label{YSDEF}}
{{\bf{J}}_1} = {\left[ {\begin{array}{*{20}{c}}
		{j_{11}}&{j_{12}}&{...}&{{j_{1m_1}}}
		\end{array}} \right]^T}.
\end{equation}
Since (\ref{SINGLEJ}) is enforced, only single equivalent electric current density ${{\bf{J}}_1}$ is required. By substituting (\ref{Y1}) and (\ref{EQUY1}) into (\ref{SINGLEJ}), ${{\bf{J}}_1}$ is obtained as
\begin{equation}{\label{YSDEF}}
\mathbf{J}_1=\mathbf{Y}_{s_1}\mathbf{E}_1,
\end{equation}
where ${\bf{Y}}_{s_1}$ is the DSAO \cite{DSAOARBSHAPEDCIM} and can be expressed as
\begin{equation}{\label{YS1}}
{\mathbf{Y}}_{s_1}={\mathbf{Y}}_1-\widehat{\mathbf{Y}}_1=\left[ {\mathbf{P}}_{1}^{\left( 1 \right)} \right] ^{-1}\left( \frac{1}{2}{\mathbf{L}}_1-{\mathbf{U}}_1^{(1)} \right) -\left[ \widehat{\mathbf{P}}_{1}^{\left( 1 \right)} \right] ^{-1}\left( \frac{1}{2}{\mathbf{L}}_1-\widehat{\mathbf{U}}_1^{(1)} \right) .
\end{equation}

\subsection{The Single-source SIE for Objects Embedded in Two Layered Media}
\begin{figure}
	\begin{minipage}[h]{0.3\linewidth}\label{FIG2A}
		\centerline{\includegraphics[scale=0.35]{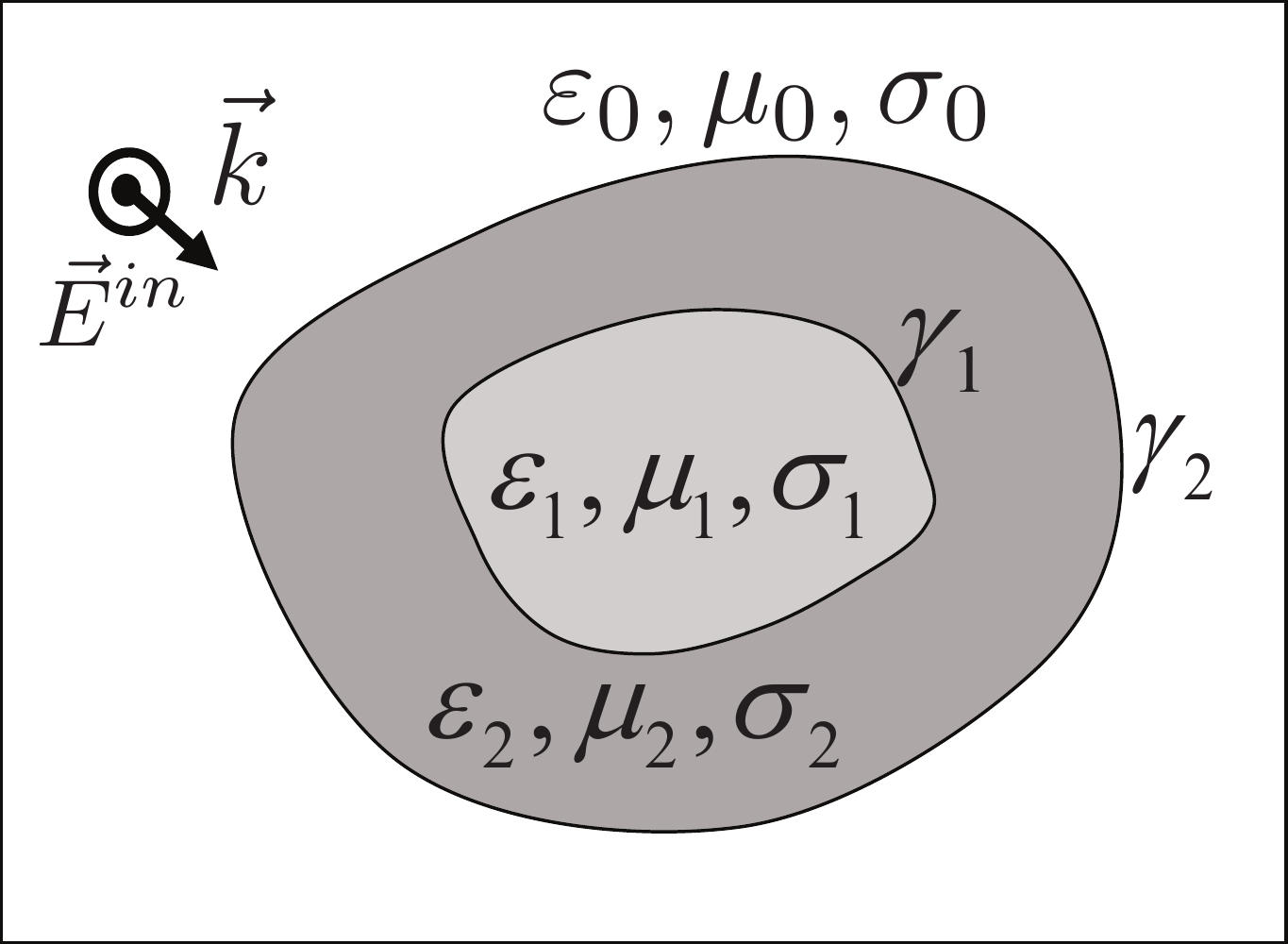}}
		\centerline{(a)}
	\end{minipage}
	\hfill
	\begin{minipage}[h]{0.3\linewidth}\label{FIG2B}
		\centerline{\includegraphics[scale=0.35]{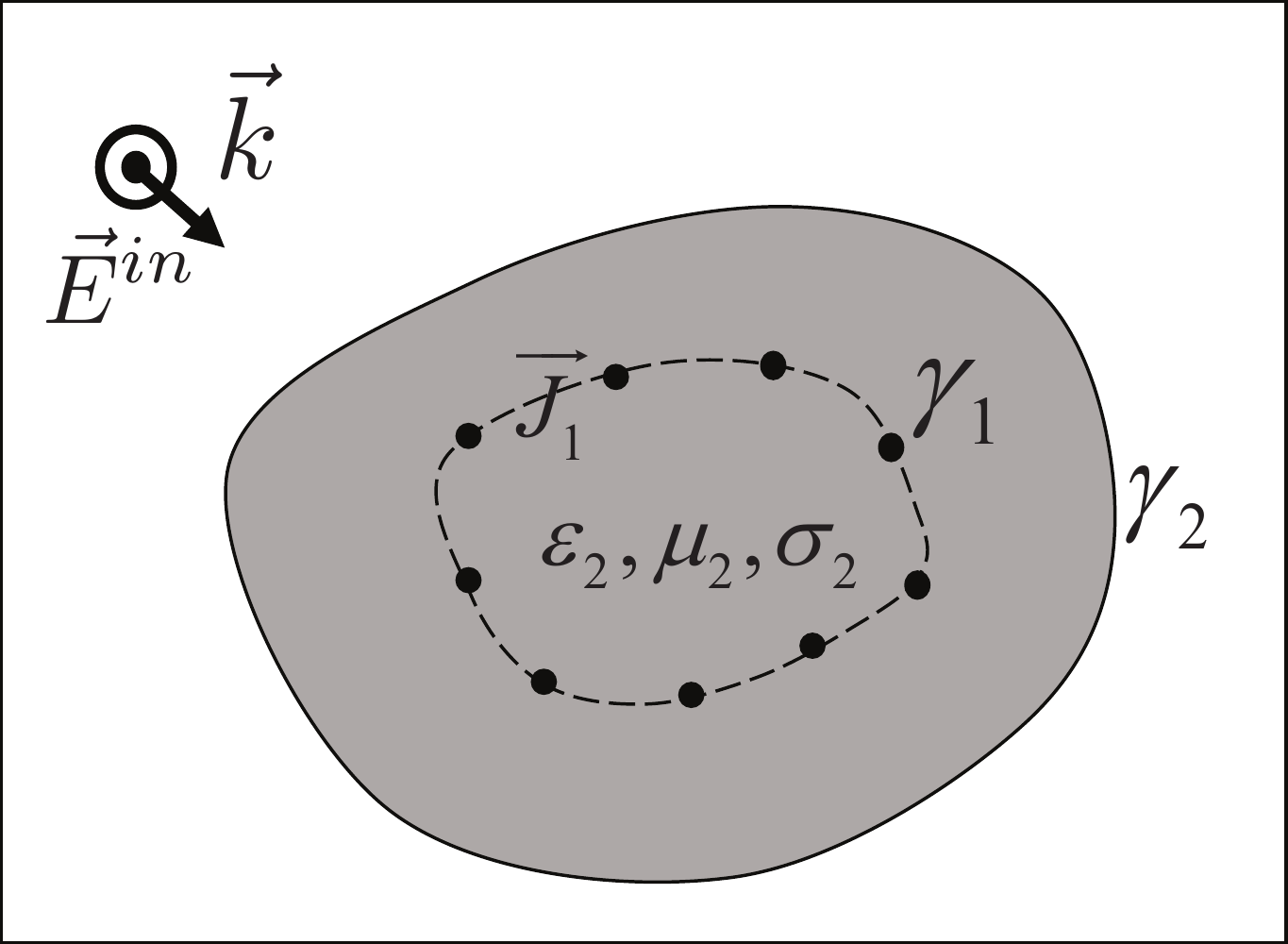}}
		\centerline{(b)}
	\end{minipage}
	\hfill
	\begin{minipage}[h]{0.3\linewidth}\label{FIG2C}
		\centerline{\includegraphics[scale=0.35]{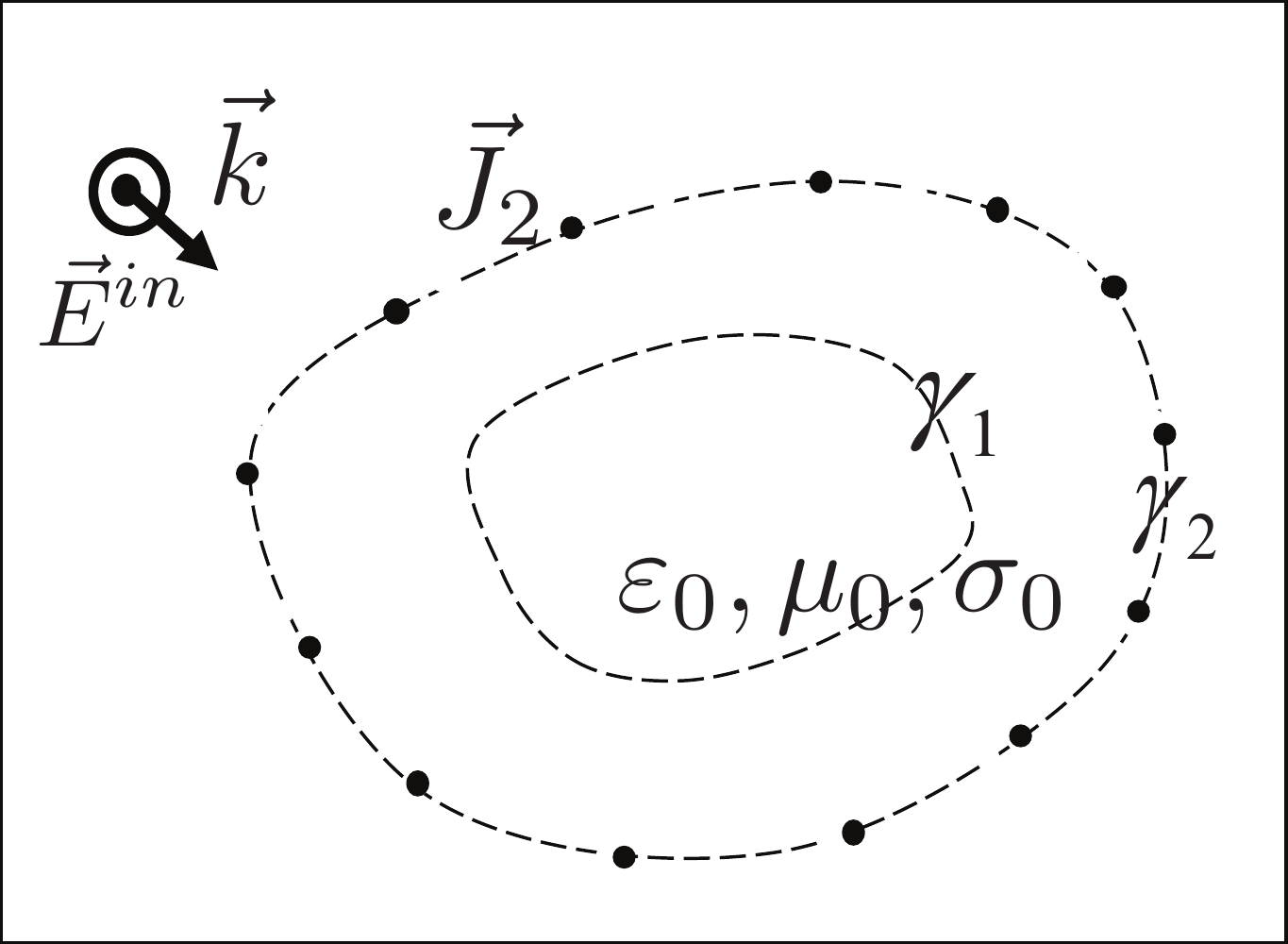}}
		\centerline{(c)}
	\end{minipage}
	\caption{(a) The original two layered model, (b) the equivalent model with the innermost medium replaced by its surrounding (the second) medium and enforcing the surface current density $\vec{J}_1$ on $\gamma_1$, and (c) the equivalent model with the inner medium replaced by its surrounding medium and enforcing the surface current density $\vec{J}_2$ on $\gamma_2$}
	\label{fig_2}
\end{figure}

\subsubsection{The original model}
We now consider a slightly more complex scenario in which a layered medium with the permittivity ${\varepsilon _2}$, the permeability ${\mu _2}$ and the conductivity ${\sigma_2}$ encloses a penetrable object as shown in Fig. 3(a). For this problem, we derive an equivalent model similar to that in previous subsection, in which objects are replaced by its surrounding medium with the permittivity ${\varepsilon _0}$, the permeability ${\mu _0}$ and the conductivity ${\sigma_0}$, and an equivalent current density ${{\vec J}_1}$ on  ${\gamma_1}$  as shown in Fig. 3(b) is enforced to ensure the fields in the outer region unchanged. By comparing with Fig.2 (a) and Fig.3 (b), it is easy to find that this problem is similar to each other. The only difference is that there is an equivalent current density ${{\vec J}_1}$, which is derived in the previous subsection, inside the new equivalent object. It should be careful to handle this current density. In $\gamma_2$, the Helmholtz equation is used to calculate electric fields in the inner region, and note that the inhomogeneous Helmholtz equation is required due to the existence of the equivalent surface current density $\vec{J}_1$ introduced in previous subsection to represent the penetrable object. Therefore, the inhomogeneous Helmholtz equation can be expressed as
\begin{equation}{\label{INHHELZ}}
{\nabla ^2}E_{2} + k_2^2E_{2} = j\omega {\mu _2}{J_1},
\end{equation}
where ${E_2}$ is the electric field inside the boundary ${\gamma_2}$, and ${J_1}$ is the equivalent current density on ${\gamma_1}$ after the first equivalent theorem applied in  Section 2.2. Through solving (\ref{INHHELZ}) with the second  scalar Green function theorem \cite{EQUIVALENT}, we obtain
\begin{equation}{\label{INTEGRAL2}}
\begin{aligned}
TE_2\left( \vec{r} \right) =\oint_{\gamma _2}{\left[ G_2\left( \vec{r},\vec{r}' \right) \frac{\partial E_2\left( \vec{r}' \right)}{\partial n'}-\frac{\partial G_2\left( \vec{r},\vec{r}' \right)}{\partial n'}E_2\left( \vec{r}' \right) \right]}dr'-\int_{\gamma _1}{j\omega \mu _2G_2J_1}ds.
\end{aligned}
\end{equation}
It should be noted that  $T=1/2$ when the source and observation points are located on ${\gamma_2}$, otherwise,   $T=1$.

We expand the electric field ${E_2}$ and the magnetic related field ${{\partial E_{2}(\vec r')}}/{{\partial n'}}$ in  (\ref{INTEGRAL2}) using the pulse basis functions and then test it using the Galerkin scheme on ${\gamma_1}$. Then, the electric field on the left side of (\ref{INTEGRAL2}) is the electric field on ${\gamma_1}$. We collect the expansion coefficient into column vectors and write it into compact form as follows
\begin{equation}{\label{TestS1}}
\mathbf{L}_1\mathbf{E}_1=\mathbf{U}_{1}^{\left( 2 \right)}\mathbf{E}_2+\mathbf{P}_{1}^{\left( 2 \right)}\mathbf{H}_2+\mathbf{G}_{1}^{\left( 2 \right)}\mathbf{J}_1.
\end{equation}
Then, we further test (\ref{INTEGRAL2}) on $\gamma_2$, and obtain the following compact form as
\begin{equation}{\label{TestS2}}
\frac{1}{2}\mathbf{L}_2\mathbf{E}_2=\mathbf{U}_{2}^{\left( 2 \right)}\mathbf{E}_2+\mathbf{P}_{2}^{\left( 2 \right)}\mathbf{H}_2+\mathbf{G}_{2}^{\left( 2 \right)}\mathbf{J}_1,
\end{equation}
where ${{\bf{E}}_1}$ and ${{\bf{E}}_2}$ denote the electric field expansion coefficients on ${\gamma_1}$ and ${\gamma_2}$, respectively, expressed as ${{\bf{E}}_1} = {\left[ {\begin{array}{*{20}{c}}
		{e_{11}}&{e_{12}}&{...}&{e_{1m_1}}
		\end{array}} \right]^T}$, ${{\bf{E}}_2} = {\left[ {\begin{array}{*{20}{c}}
		{e_{21}}&{e_{22}}&{...}&{e_{2m_2}}
		\end{array}} \right]^T}$, and ${{\bf{H}}_2}$ is the magnetic field expansion coefficients on ${\gamma_2}$ expressed as ${{\bf{H}}_2} = {\left[ {\begin{array}{*{20}{c}}
		{h_{21}}&{h_{22}}&{...}&{h_{2m_2}}
		\end{array}} \right]^T}$. $\mathbf{L}_1$ and $\mathbf{L}_2$ are diagonal matrixes with the length of each segments as their entries on $\gamma_1$ and $\gamma_2$, respectively. ${{\bf{J}}_1}$ is the surface equivalent current density expansion vector, defined in (\ref{YSDEF}), on ${\gamma_1}$ obtained in Section 2.2. The elements of ${{{\bf{U}}^{(1)}_2}}$, ${{{\bf{U}}^{(2)}_2}}$, ${{\bf{P}}^{(2)}_1}$, ${{\bf{P}}^{(2)}_2}$ ${{\bf{G}}^{(2)}_1}$ and ${{\bf{G}}^{(2)}_2}$ are expressed as
\begin{equation}{\label{U12}}
\left[ \mathbf{U}_{1}^{\left( 2 \right)} \right] _{m,n}=\int_{\gamma _{1_m}}{\int_{\gamma _{2_n}}{k_2\frac{\vec{d}_m\cdot \hat{n}'}{d_m}}G_{2}^{'}\left( \vec{r},\vec{r}' \right) dr'}dr,
\end{equation}
\begin{equation}{\label{U2}}
\left[ \mathbf{U}_{2}^{\left( 2 \right)} \right] _{m,n}=\int_{\gamma _{2_m}}{\int_{\gamma _{2_n}}{k_2\frac{\vec{d}_m\cdot \hat{n}'}{d_m}}G_{2}^{'}\left( \vec{r},\vec{r}' \right) dr'}dr,
\end{equation}
\begin{equation}{\label{P12}}
\left[ \mathbf{P}_{1}^{\left( 2 \right)} \right] _{m,n}=\int_{\gamma _{1_m}}{\int_{\gamma _{2_n}}{\omega _2\mu _2jG_2\left( \vec{r},\vec{r}' \right) dr'}}dr,
\end{equation}
\begin{equation}{\label{P2}}
\left[ \mathbf{P}_{2}^{\left( 2 \right)} \right] _{m,n}=\int_{\gamma _{2_m}}{\int_{\gamma _{2_n}}{\omega _2\mu _2jG_2\left( \vec{r},\vec{r}' \right) dr'}}dr,
\end{equation}
\begin{equation}{\label{G12}}
\left[ \mathbf{G}_{1}^{\left( 2 \right)} \right] _{m,n}=-\int_{\gamma _{1_m}}{\int_{\gamma _{1_n}}{\omega _2\mu _2jG_2\left( \vec{r},\vec{r}' \right) dr'}}dr,
\end{equation}
\begin{equation}{\label{G2}}
\left[ \mathbf{G}_{2}^{\left( 2 \right)} \right] _{m,n}=-\int_{\gamma _{2_m}}{\int_{\gamma _{1_n}}{\omega _2\mu _2jG_2\left( \vec{r},\vec{r}' \right) dr'}}dr,
\end{equation}
where ${{\vec d}_m} = \vec r' - {{\vec r}}$, ${d_m} = \left| {\vec r' - {{\vec r}}} \right|$, $G_2=-\frac{j}{4}H_{0}^{\left( 2 \right)}\left( k_2\rho \right)$, $G_{2}^{'}=-\frac{j}{4}H_{1}^{\left( 2 \right)}\left( k_2\rho \right)$.

After substituting  (\ref{YSDEF}) into  (\ref{TestS1}) and making some mathematical manipulations, ${{\bf{E}}_1}$ is expressed in terms of ${{\bf{E}}_2}$ and  ${{\bf{H}}_2}$ as
\begin{equation}{\label{TransS1}}
\mathbf{E}_1=\underset{\mathbf{V}_{1}^{\left( 2 \right)}}{\underbrace{\left( \mathbf{L}_1-\mathbf{G}_{1}^{\left( 2 \right)}\mathbf{Y}_{s_1} \right) ^{-1}}}\left( \mathbf{U}_{1}^{\left( 2 \right)}\mathbf{E}_2+\mathbf{P}_{1}^{\left( 2 \right)}\mathbf{H}_2 \right) .
\end{equation}
Then, after substituting (\ref{YSDEF}) and (\ref{TransS1}) into (\ref{TestS2}), the relationship of ${{\bf{E}}_2}$ and ${{\bf{H}}_2}$ on ${\gamma_2}$ is obtained as
\begin{equation}{\label{Y2}}
\begin{aligned}
{{\bf{H}}_2} = {{\bf{Y}}_{2}}{{\bf{E}}_2}.
\end{aligned}
\end{equation}
where
\begin{equation}{\label{Y2}}
\begin{aligned}
\mathbf{Y}_2=\left[ \mathbf{P}_{2}^{\left( 2 \right)}+\mathbf{F}_{2}^{\left( 2 \right)}\mathbf{P}_{1}^{\left( 2 \right)} \right] ^{-1}\left( \frac{1}{2}\mathbf{L}_2-\mathbf{U}_{2}^{\left( 2 \right)}-\mathbf{F}_{2}^{\left( 2 \right)}\mathbf{U}_{1}^{\left( 2 \right)} \right) ,
\end{aligned}
\end{equation}
where $\mathbf{F}_{2}^{\left( 2 \right)}=\mathbf{G}_{2}^{\left( 2 \right)}\mathbf{Y}_{s_1}\mathbf{V}_{1}^{\left( 2 \right)}$.

\subsubsection{The equivalent model for two layered objects}
According to the equivalence theorem, the equivalent model is shown in Fig. 3(c), where the inner medium is replaced by its surrounding medium with parameters ${\varepsilon _0},{\mu _0},{\sigma_0}$. In the equivalent model, there is no current sources existing inside the equivalent object. However, another surface equivalent current density, ${\vec J_2}$, on the boundary ${\gamma_2}$ is introduced to enforce the fields exactly the same as those in the original model.

With the similar procedure in Section 2.2 for the equivalent model of a penetrable object, the relationship between the electric field ${{\bf{E}}_2}$ and the magnetic field ${\widehat{\bf{H}}_2}$ on ${\gamma_2}$ in equivalent model is obtained as
\begin{equation}{\label{Y22}}
\widehat{\mathbf{H}}_2=\underset{\widehat{\mathbf{Y}}_2}{\underbrace{\left[ \widehat{\mathbf{P}}_{2}^{\left( 2 \right)} \right] ^{-1}\left( \frac{1}{2}\mathbf{L}_2-\widehat{\mathbf{U}}_{2}^{\left( 2 \right)} \right) }}\mathbf{E}_2.
\end{equation}

After substituting (\ref{Y2}) and (\ref{Y22}) into (\ref{SINGLEJ}), we can obtain the equivalent current density ${\widehat {\bf{J}}_{2}}$ on ${\gamma_2}$ as
\begin{equation}
{{\bf{J}}_{2}} = {{\bf{H}}_2} - {{{\bf{\widehat H}}}_2} = \underbrace{({\bf{Y}}_2 - {\widehat {\bf{Y}}_2})}_{{\bf{Y}}_{s_2}}{{\bf{E}}_2}.
\end{equation}

As shown in Fig. 3(c), the objects embedded in multilayers are replaced by its background medium and only a single surface electric current density is introduced on $\gamma_2$. The PMCHWT formulation has unknowns on both $\gamma_1$ and $\gamma_2$. However, the proposed approach only has unknowns residing on the outermost boundary $\gamma_2$, which will greatly reduce the overall number of unknowns. To derive a unified approach for any objects embedded in multilayers, we further generalize the proposed approach in the next subsection.

\subsection{Generalization of the Proposed Approach to Arbitrary Objects Embedded in Multilayers}
\begin{figure}
	\begin{minipage}[h]{0.5\linewidth}\label{FIG2A}
		\centerline{\includegraphics[scale=0.4]{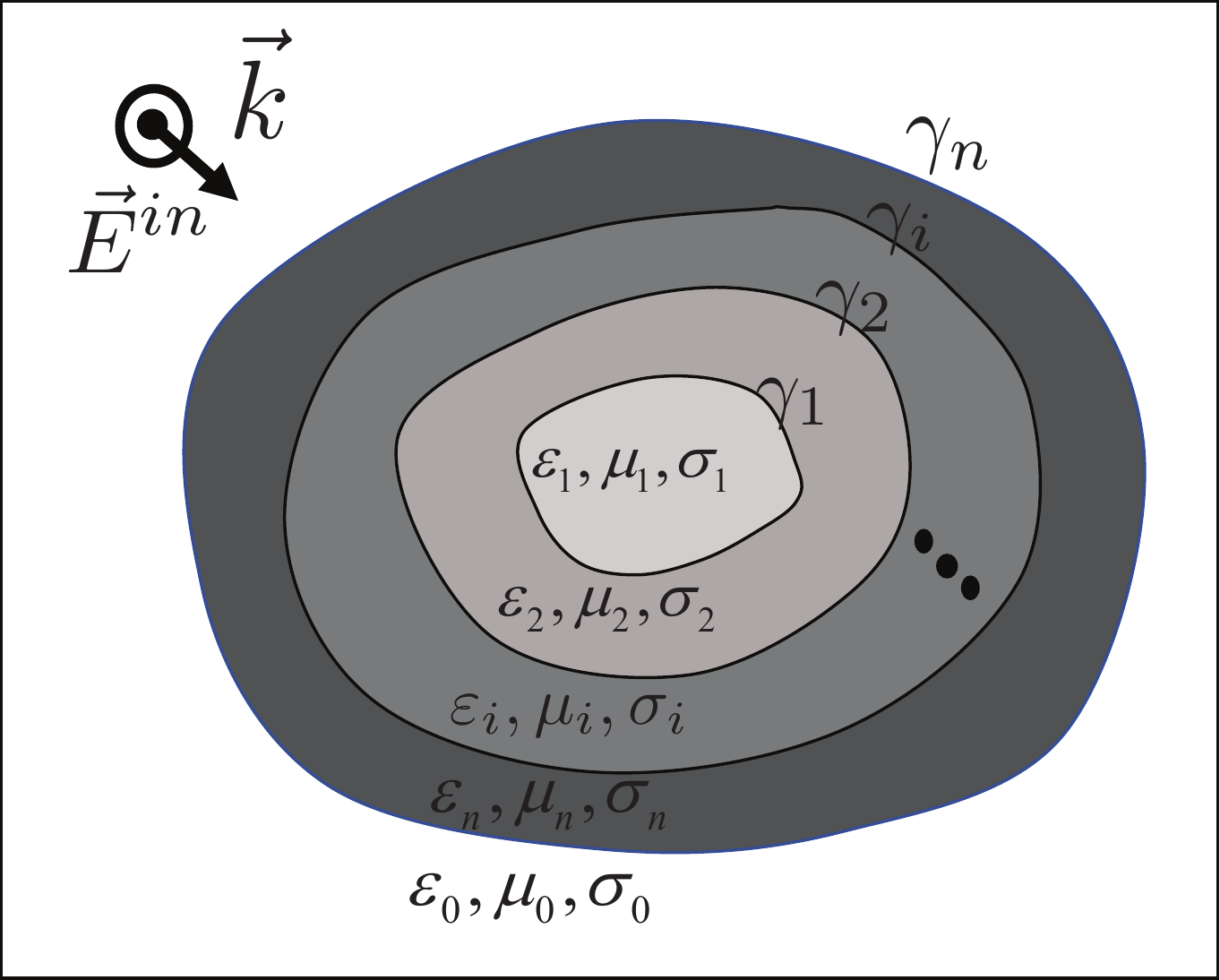}}
		\centerline{(a)}
	\end{minipage}
	\hfill
	\begin{minipage}[h]{0.5\linewidth}\label{FIG2B}
		\centerline{\includegraphics[scale=0.4]{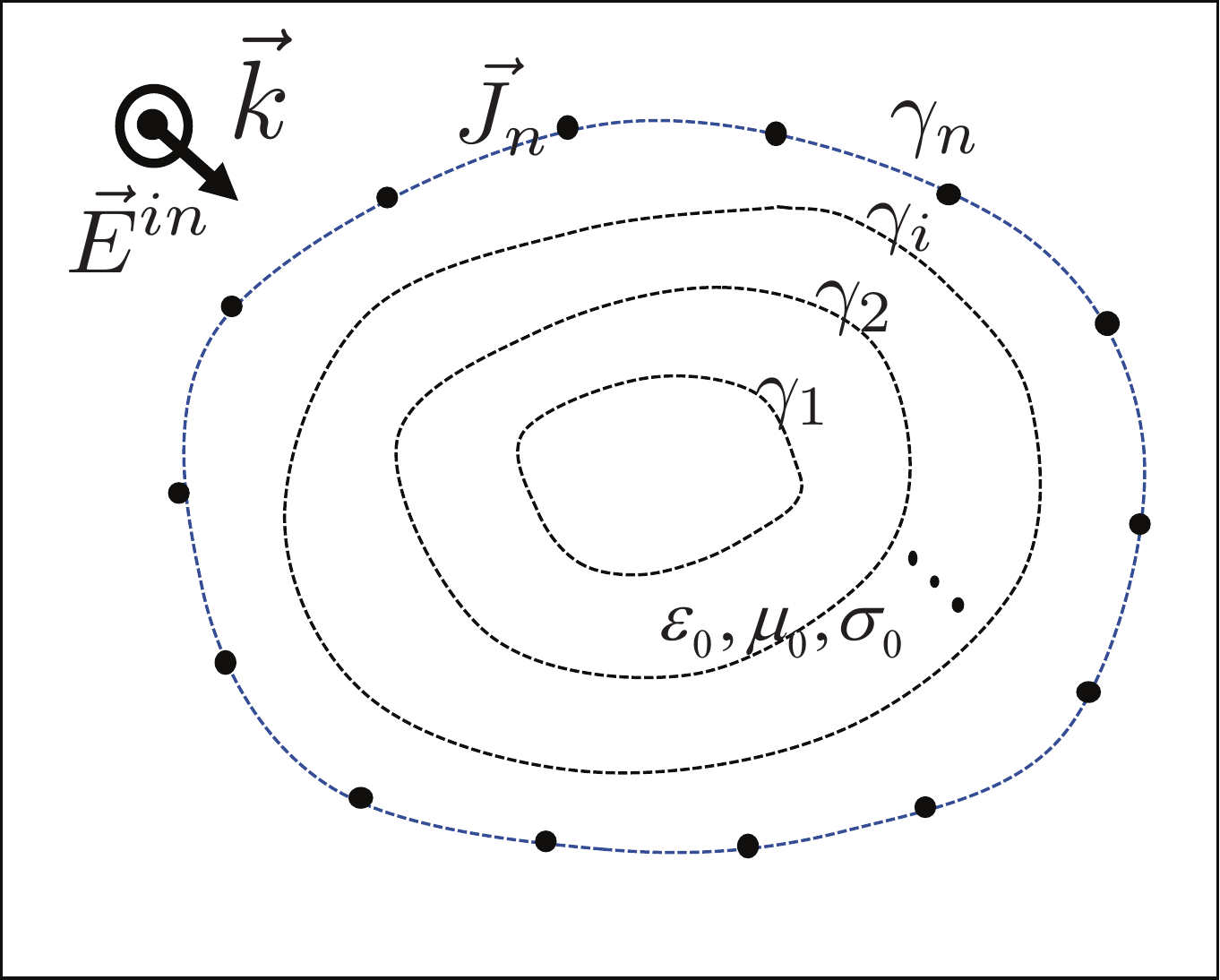}}
		\centerline{(b)}
	\end{minipage}
	\caption{(a) The original object embedded into multilayer dielectric media and (b) the equivalent model with the inner medium replaced by its background medium and enforcing the surface electric current density $\vec J_n$ on the outermost boundary $\gamma_n$}
	\label{fig_2}
\end{figure}
Once we have presented the approach to model the two layered embedded objects as shown in the previous two subsections, it is straightforward to extend the approach to model any objects embedded in multilayers.
Similarly, for objects embedded a $n$ layered medium, we can apply the proposed approach recursively and obtain the surface equivalent current on the outmost boundary.  The first equivalent procedure is similar to that presented in the Section 2.2, and the subsequent equivalent procedure is similar to that in the Section 2.3. The surface equivalent current density induced from the $(i-1)$th medium is expressed as
\begin{equation}{\label{YSI1}}
{\mathbf{J}}_{i-1}=\mathbf{Y}_{s_{i-1}}\mathbf{E}_{i-1}.
\end{equation}
For the original object with the $i$th ($i \ge 2$) layer media, it is required to test (\ref{INTEGRAL2}) on the $(i-1)$th layer as
\begin{equation}{\label{TESTSI}}
\mathbf{L}_{i-1}\mathbf{E}_{i-1}=\mathbf{U}_{i-1}^{\left( i \right)}\mathbf{E}_i+\mathbf{P}_{i-1}^{\left( i \right)}\mathbf{H}_i+\mathbf{G}_{i-1}^{\left( i \right)}\mathbf{J}_{i-1},
\end{equation}
and on the boundary of the $i$th layer as
\begin{equation}{\label{TESTSI1}}
\frac{1}{2}\mathbf{L}_i\mathbf{E}_i=\mathbf{U}_{i}^{\left( i \right)}\mathbf{E}_i+\mathbf{P}_{i}^{\left( i \right)}\mathbf{H}_i+\mathbf{G}_{i}^{\left( i \right)}\mathbf{J}_{i-1},
\end{equation}
where ${{\bf{E}}_i}$, ${{\bf{H}}_i}$ denote the electric field on $\gamma_i$ and the entities of  other three matrices are denoted as follows
\begin{equation}
\left[ \mathbf{U}_{p}^{\left( q \right)} \right] _{m,n}=\int_{\gamma _{p_m}}{\int_{\gamma _{q_n}}{k_q\frac{\vec{d}_m\cdot \hat{n}'}{d_m}G_{q}^{}\left( \vec{r},\vec{r}' \right)}dr'}dr,
\end{equation}
\begin{equation}
\left[ \mathbf{P}_{p}^{\left( q \right)} \right] _{m,n}=\int_{\gamma _{p_m}}{\int_{\gamma _{q_n}}{\omega _q\mu _qG_q\left( \vec{r},\vec{r}' \right)}dr'}dr,
\end{equation}
\begin{equation}
\left[ \mathbf{G}_{p}^{\left( q \right)} \right] _{m,n}=-\int_{\gamma _{p_m}}{\int_{\gamma _{{(q-1)}_n}}{\omega _q\mu _qG_q\left( \vec{r},\vec{r}' \right)}dr'}dr.
\end{equation}

Through some mathematical manipulations using  (\ref{YSI1}),  (\ref{TESTSI}) and  (\ref{TESTSI1}), we obtain
\begin{equation}
\begin{aligned}
\mathbf{H}_i=\left[ \mathbf{P}_{i}^{\left( i \right)}+\mathbf{F}_{i}^{\left( i \right)}\mathbf{P}_{i-1}^{\left( i \right)} \right] ^{-1}\left( \frac{1}{2}\mathbf{L}_i-\mathbf{U}_{i}^{\left( i \right)}-\mathbf{F}_{i}^{\left( i \right)}\mathbf{U}_{i-1}^{\left( i \right)} \right)\mathbf{E}_i,
\end{aligned}
\end{equation}
where $\mathbf{F}_{i}^{\left( i \right)}=\mathbf{G}_{i}^{\left( i \right)}\mathbf{Y}_{s_{i-1}}\mathbf{V}_{i-1}^{\left( i \right)}$, $\mathbf{V}_{i-1}^{\left( i \right)}=\left( \mathbf{L}_{i-1}-\mathbf{G}_{i-1}^{\left( i \right)}\mathbf{Y}_{s_{i-1}} \right) ^{-1}$. Then, it can be rewritten into a more compact form as
\begin{equation}\label{GYSI}
{{\bf{H}}_i} = {{\bf{Y}}_i}{{\bf{E}}_i}.
\end{equation}

For the equivalent model, the procedure is similar to that in the equivalent model as shown in Section 2.3. The surface discretized magnetic and electric fields can be expanded as
\begin{equation}
\widehat{\mathbf{H}}_i=\underset{\widehat{\mathbf{Y}}_i}{\underbrace{\left[ \widehat{\mathbf{P}}_{i}^{\left( i \right)} \right] ^{-1}\left( \frac{1}{2}\mathbf{L}_i-\widehat{\mathbf{U}}_{i}^{\left( i \right)} \right) }}\mathbf{E}_i.
\end{equation}
Therefore, the equivalent current density ${{\bf{J}}_{i}}$ on $\gamma_i$ is expressed as
\begin{equation}{\label{YSI}}
{{\bf{J}}_{i}} = {{\bf{H}}_i} - {{{\bf{\widehat H}}}_i} = \underbrace{({\bf{Y}}_i - {\widehat {\bf{Y}}_i})}_{{\bf{Y}}_{s_i}}{{\bf{E}}_i}.
\end{equation}
Then, the equivalent model obtained is shown in Fig. 4(b), and the single surface equivalent current ${{\bf{J}} _n} = {\bf{Y}}_{s_n}{{\bf{E}}_n}$ on the outermost layer boundary ${\gamma_n}$ is obtained. The original object is replaced by its surrounding medium along with a surface equivalent current density ${{\bf{J}} _n}$ on $\gamma_n$ as shown in Fig. 4(b).

\subsection{Multiple Scattering Objects}
When there are multiple scattering objects involved on $\gamma_i$, we first compute the equivalent current density for each object according to the proposed approach in Section 2.4 as
\begin{equation}
{\bf{J}}_{i}^{(p)} = {\bf{Y}}_{s_i}^{(p)}{\bf{E}}_{i}^{(p)},
\end{equation}
where the superscript $(p)$ represents the $p$th object, ${\bf{J}}_{i}^{(p)}$, ${\bf{E}}_{i}^{(p)}$ denote the electric current density and electric field of the $p$th object, respectively.

Then, we collect all the surface equivalent current together and obtain
\begin{equation}
{{\bf{J}}_i} = {{\bf{Y}}_{s_i}}{{\bf{E}}_i},
\end{equation}
where the surface admittance operator ${{\bf{Y}}_{s_i}}$ of the $i$th layer is a diagonal block matrix assembling from the surface equivalent operator ${\bf{Y}}_{s_i}^{(p)}$ of each object and can be expressed as
\begin{equation}
{{\bf{Y}}_{s_i}} = \left[ {\begin{array}{*{20}{c}}
	{{\bf{Y}}_{s_i}^{(1)}}&{}&{}&{}\\
	{}&{{\bf{Y}}_{s_i}^{(2)}}&{}&{}\\
	{}&{}&{...}&{}\\
	{}&{}&{}&{{\bf{Y}}_{s_i}^{(p)}}
	\end{array}} \right].
\end{equation}
${{\bf{E}}_i}$ is a column vector assembling all electric field coefficients of each object and is expressed  as
\begin{equation}
\begin{aligned}
{{\bf{E}}_i} = {\left[ {\begin{array}{*{20}{c}}
		{E_1^{(1)}}&{...}&{E_{m_1}^{(1)}}&{...}&{E_1^{(p)}}&{...}&{E_{m_p}^{(p)}}\end{array}} \right]^T},
\end{aligned}
\end{equation}
and ${{\bf{J}}_i}$ is expressed as
\begin{equation}
{{\bf{J}}_i} = {\left[ {\begin{array}{*{20}{c}}
		{J_1^{(1)}}&{...}&{J_{m_1}^{(1)}}&{...}&{J_1^{(p)}}&{...}&{J_{m_p}^{(p)}}
		\end{array}} \right]^T}.
\end{equation}

\subsection{Scattering Modeling of the Exterior Problem}
The electric field $\vec{E}$ outside the object is the superposition of the incident field ${{\vec{E}}^i}$ and the scattered field ${{\vec{E}}^s}$.  Since the surface equivalent current ${\vec{J}}_{n}$ exists on the outermost layer of the equivalent object, the induced electric field and magnetic field by ${\vec{J}}_{n}$ are expressed as
\begin{equation}{\label{SUPERPE}}
E^s\left( \vec{r} \right) =-j\omega \mu \int_{\gamma _n}{J_n\left( \vec{r}' \right)}G_0\left( \vec{r},\vec{r}' \right) ds',
\end{equation}
where $\vec r'$ on the outermost boundary ${\gamma_n}$ of the object, ${G_0}$ is the Green function expressed as ${G_0} =  -
j H_0^{(2)}({k_0}\rho )/4$, where ${k_0}$ is the wavenumber in the background medium. Obviously, ${{J}}_n$ is the equivalent current on ${\gamma_n}$.
Therefore, the electric and magnetic fields can be simplified as
\begin{equation}{\label{EFIE}}
\mathbf{E}_n=\mathbf{G}_{n}^{\left( n+1 \right)}\mathbf{J}_n+\mathbf{E}_n^i,
\end{equation}
where the elements in $\mathbf{E}$ and $\mathbf{E}^i$ are the integral of the total electric field and the incident electric field over each segment, respectively.

We substitute (\ref{YSI}) into (\ref{EFIE}) and obtain the electric field ${{\bf{E}}_n}$ on the outermost boundary ${\gamma_n}$ as
\begin{equation}
\begin{aligned}
\mathbf{E}_n=\left( \mathbf{I}+\mathbf{G}_{n}^{\left( n+1 \right)}\mathbf{Y}_s \right) ^{-1}\mathbf{E}_n^i,
\end{aligned}
\end{equation}
where $\bf{I}$ is an identity matrix. Once the electric field  ${\bf{E}}_n$ on $\gamma_n$ is obtain, we can easily calculate the surface equivalent current ${\bf{J}}_n$ and other interested parameters.

\section{Two Special Cases}
\subsection{PEC Embedded Objects}

\begin{figure}
	\begin{minipage}[h]{0.5\linewidth}\label{FIG5B}
		\centerline{\includegraphics[scale=0.4]{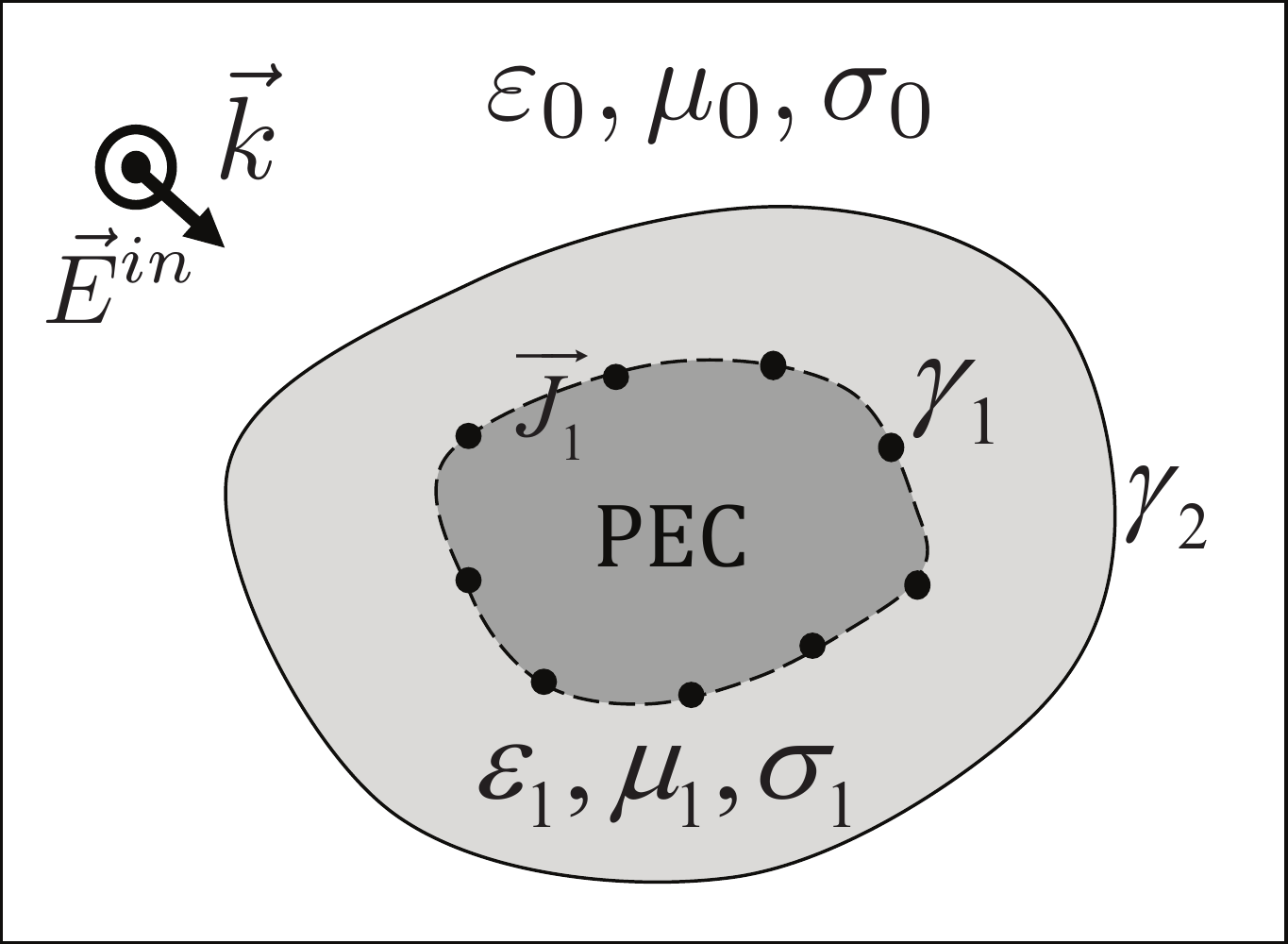}}
		\centerline{(a)}
	\end{minipage}
	\hfill
	\begin{minipage}[h]{0.5\linewidth}\label{FIG5C}
		\centerline{\includegraphics[scale=0.4]{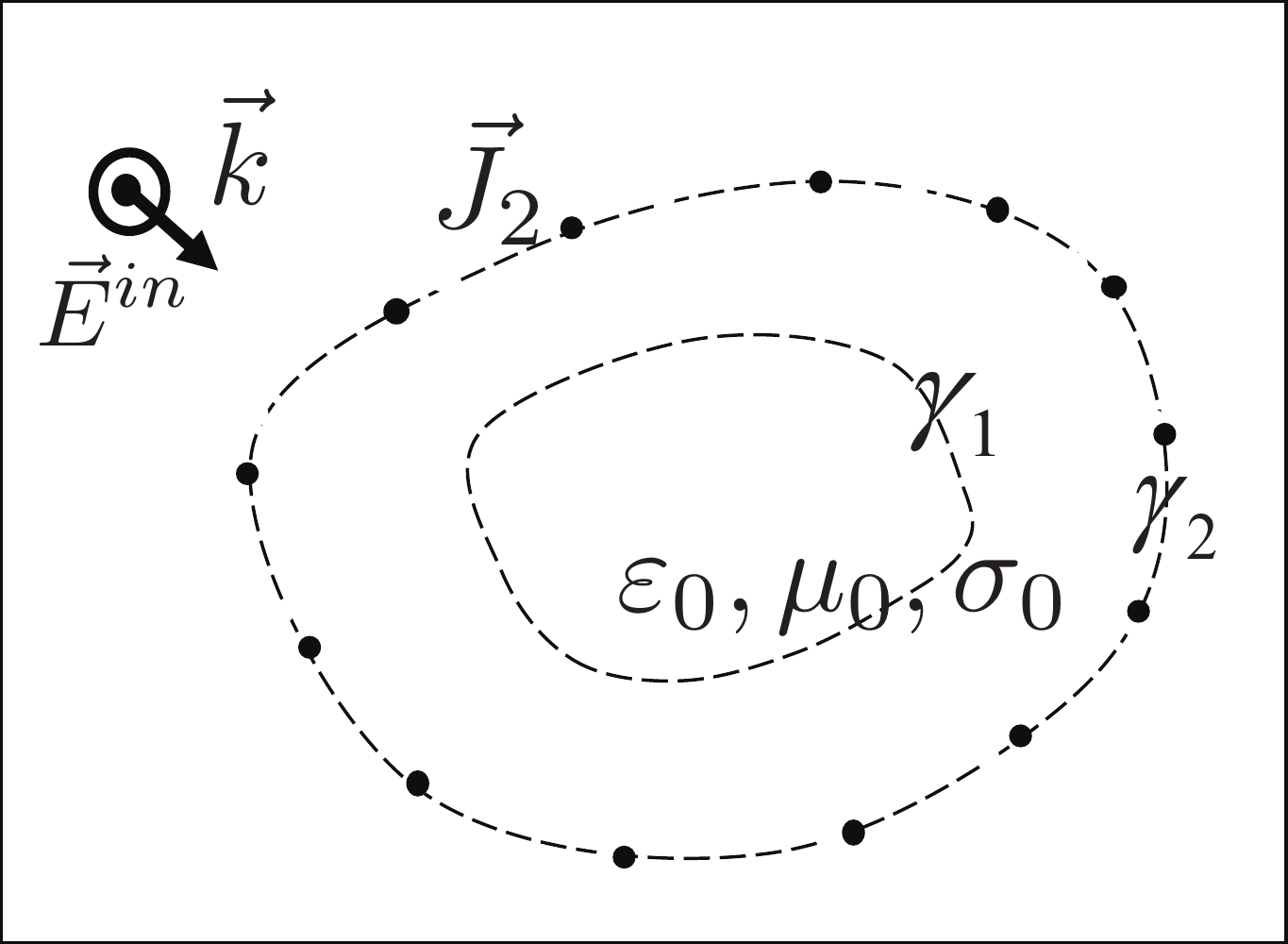}}
		\centerline{(b)}
	\end{minipage}
	\caption{(a) Original PEC embedded object and (b) the equivalent model with the inner medium replaced by its surrounding medium and enforcing the surface currents density on $\gamma_2$}
	\label{fig_5}
\end{figure}

In this section, we derive the formulation in the proposed approach for the PEC objects embedded multilayers.
\subsubsection{A single PEC object}
For a single PEC object, there is a current density $\vec{J}_1$ on $\gamma_1$ as shown in Fig. 5(a) and fields inside $\gamma_1$ vanish. This scenario is different from that present in the Section 2.2. Therefore, for a single PEC object, we do not need to apply the equivalence theorem to obtain the surface equivalent current on its boundary. The current $\vec{J}_1$ is the true surface current density flowing on $\gamma_1$.
\subsubsection{A PEC object with a layer of medium}
The PEC object with a layer of medium is shown in Fig. 5(a). It can be seen from Section 3.1 that the object can be equivalent to Fig. 5(b). The permittivity, the permeability and the conductivity of medium inside $\gamma_2$ are ${\varepsilon _1}$, ${\mu _1}$ and ${\sigma_1}$, and the current $\vec{J}_1$ exists on $\gamma_1$.

According to the inhomogeneous Helmholtz equation, we can solve the electric field on $\gamma_2$ as
\begin{equation}{\label{PECIE}}
\begin{aligned}
TE_{2}(\vec r) = \oint_{{\gamma_2}} {\left[ {G_2(\vec r,\vec r')\frac{{\partial E_{2}(\vec r')}}{{\partial n'}} - \frac{{\partial G_2(\vec r,\vec r')}}{{\partial n'}}E_{2}(\vec r')} \right]} dr'- \int_{\gamma_1} { j\omega {\mu _2}G_2{J_1}} ds.
\end{aligned}
\end{equation}
Then, we test the above equation on $\gamma_1$. Since the electric field vanishes on $\gamma_1$,  we obtain the following formulation
\begin{equation}{\label{S1PEC}}
{{\bf{0}}} = {\bf{U}}_1^{(2)}{{\bf{E}}_2} + {\bf{P}}_1^{(2)}{{\bf{H}}_2} + {\bf{G}}_1^{(2)}{{\bf{J}}_1}.
\end{equation}
Next, we test (\ref{PECIE}) on $\gamma_2$ and obtain
\begin{equation}{\label{S2PEC}}
\frac{1}{2}\mathbf{L}_2\mathbf{E}_2=\mathbf{U}_{2}^{\left( 2 \right)}\mathbf{E}_2+\mathbf{P}_{2}^{\left( 2 \right)}\mathbf{H}_2+\mathbf{G}_{2}^{\left( 2 \right)}\mathbf{J}_1.
\end{equation}
The entries of each matrix are the same as those in (\ref{U12})-(\ref{G2}).

According to the Section 2.3, by using (\ref{S1PEC}) and  (\ref{S2PEC}) along with some mathematical manipulations, we can obtain the relationship between the equivalent current density ${\bf{J}}_2$ on $\gamma_2$ and ${\bf{E}}_2$, which can be expressed as
\begin{equation}
{{\bf{J}}_2} = {{\bf{Y}}_{s_2}}{{\bf{E}}_2}.
\end{equation}
where
\begin{equation}
\begin{split}
\mathbf{Y}_{s_2}=\left[ \mathbf{P}_{2}^{\left( 2 \right)}-\mathbf{G}_{2}^{\left( 2 \right)}\left( \mathbf{G}_{1}^{\left( 2 \right)} \right) ^{-1}\mathbf{P}_{1}^{\left( 2 \right)} \right] ^{-1}\left[ \frac{1}{2}\mathbf{L}_2-\mathbf{U}_{2}^{\left( 2 \right)}+\mathbf{G}_{2}^{\left( 2 \right)}\left[ \mathbf{G}_{1}^{\left( 2 \right)} \right] ^{-1}\mathbf{U}_{1}^{\left( 2 \right)} \right] -\left[ \widehat{\mathbf{P}}_{2}^{\left( 2 \right)} \right] ^{-1}\left( \frac{1}{2}\mathbf{L}_2-\widehat{\mathbf{U}}_{2}^{\left( 2 \right)} \right).
\end{split}
\end{equation}

\subsubsection{A PEC object embedded into multilayered media}
When the PEC object is embedded in multiple layered media,  we can obtain the equivalent current $\vec{J}_2$ on the boundary $\gamma_2$ through Section 2.4. Then, according to the equivalence theorem and the method proposed in Section 2.4, we can derive the equivalent current $\vec{J}_n$ on $\gamma_n$ layer by layer and then obtain the final equivalent model.

\subsection{Surface Extension in the Same Medium}{\label{extension}}
\begin{figure}
	\begin{minipage}[h]{0.5\linewidth}\label{FIG5B}
		\centerline{\includegraphics[scale=0.35]{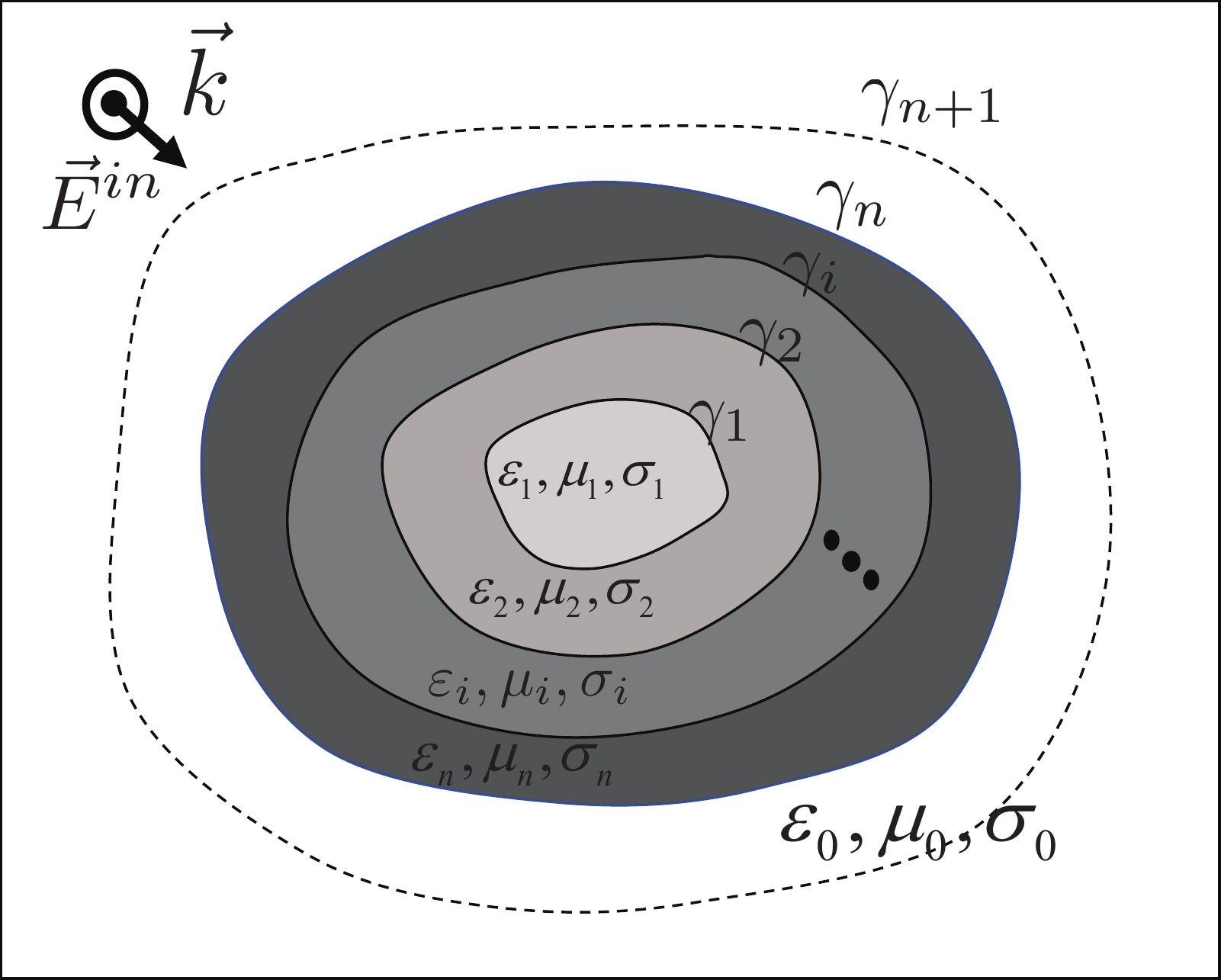}}
		\centerline{(a)}
	\end{minipage}
	\hfill
	\begin{minipage}[h]{0.5\linewidth}\label{FIG5C}
		\centerline{\includegraphics[scale=0.35]{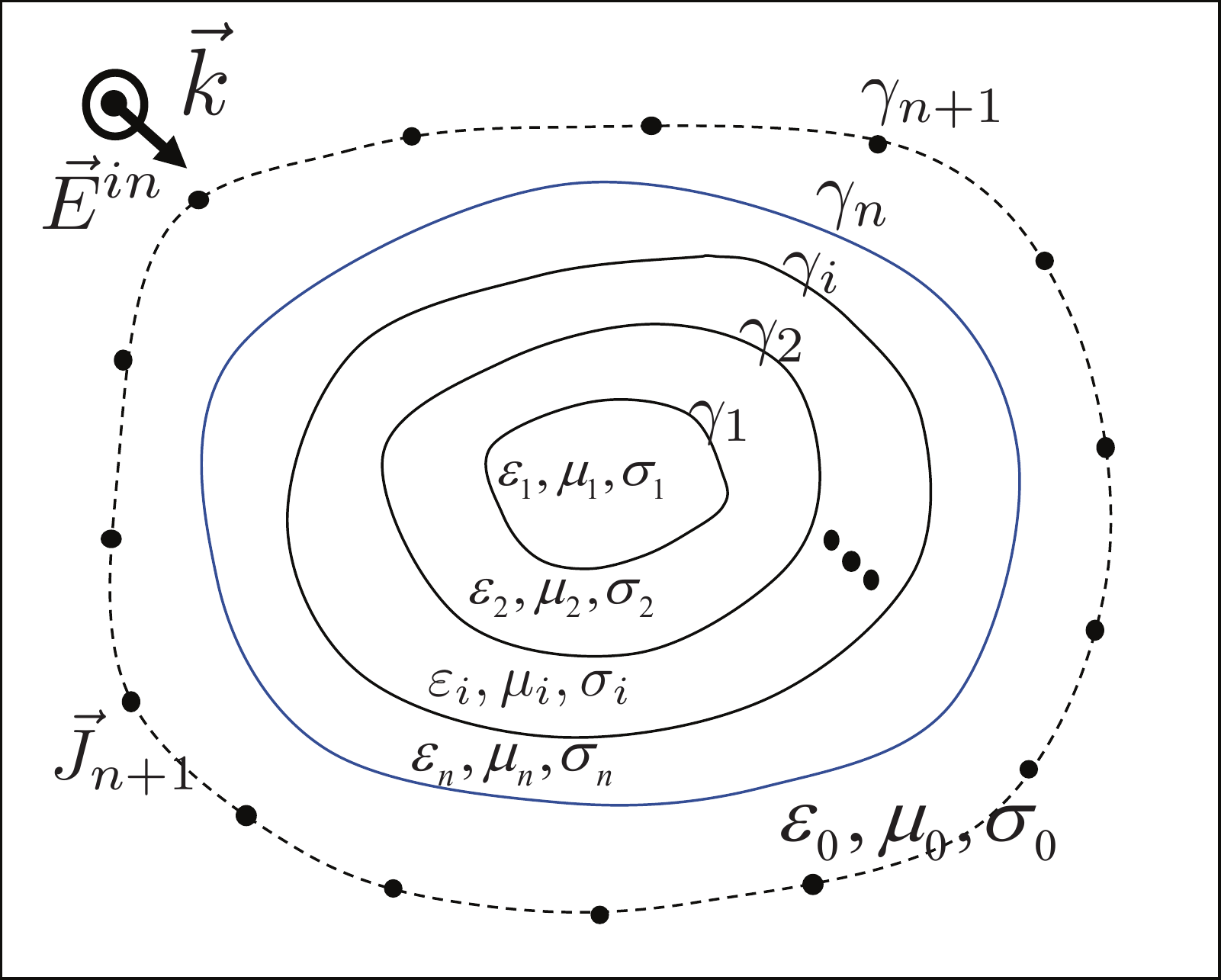}}
		\centerline{(b)}
	\end{minipage}
	\caption{(a) The original model and (b) the equivalent model with the medium replaced by its background medium and enforcing the surface current $\vec J_{n+1}$ at $\gamma_{n+1}$}
	\label{fig_2}
\end{figure}

We further extend the proposed approach to model the fictitious extension boundary in the same medium. According to the method proposed in the Section 2.4, we have obtained the equivalent current on the outermost boundary of a object embedded in multilayers as
\begin{equation}
{{\bf{J}}_n} = {{\bf{Y}}_{s_n}}{{\bf{E}}_n},
\end{equation}
and the relationship between electric and magnetic fields on the outermost boundary can be expressed as
\begin{equation}
{{\bf{H}}_n} = {{\bf{Y}}_n}{{\bf{E}}_n}.
\end{equation}

We can make a fictitious boundary $\gamma_{n+1}$ in the background medium as shown in Fig. 6(a). Then, an equivalent model is obtained through enforcing an equivalent surface current at $\gamma_{n+1}$. If we treat the medium between the target boundary $\gamma_{n+1}$ and the outermost boundary $\gamma_n$ of the object as a new one with the same constitutive parameters as those of the background medium, the proposed approach can be further extended to handle this objects.

The method in the section 2.4 is used to conduct the equivalence of a new layer of medium, as shown in the Fig. 6(b). We can obtain the relationship between tangential magnetic field and normal electric field on $\gamma_{n+1}$ of the original model as
\begin{equation}
{{\bf{H}}_{n + 1}} = {{\bf{Y}}_{n + 1}}{{\bf{E}}_{n + 1}},
\end{equation}
where
\begin{equation}
\begin{aligned}
\mathbf{Y}_{n + 1}=\left[ \mathbf{P}_{n + 1}^{\left( n + 1 \right)}+\mathbf{F}_{n + 1}^{\left( n + 1 \right)}\mathbf{P}_{n}^{\left( n + 1 \right)} \right] ^{-1}\left( \frac{1}{2}\mathbf{L}_{n+1}-\mathbf{U}_{n+1}^{\left( n+1 \right)}-\mathbf{F}_{n+1}^{\left( n+1 \right)}\mathbf{U}_{n}^{\left( n+1 \right)} \right) .
\end{aligned}
\end{equation}
Then, in the equivalent model, an surface equivalent current is enforced at $\gamma_{n+1}$ as shown in Fig. 6(b), where the relationship between tangential magnetic field and electric field on boundary $\gamma_{n+1}$ can expressed as
\begin{equation}
\widehat{\mathbf{H}}_{n+1}=\left[ \widehat{\mathbf{P}}_{n+1}^{\left( n+1 \right)} \right] ^{-1}\left( \frac{1}{2}\mathbf{L}_{n+1}-\widehat{\mathbf{U}}_{n+1}^{\left( n+1 \right)} \right) \mathbf{E}_{n+1}.
\end{equation}

For the $(n+1)$th layered medium, since the outermost medium is equal to the background medium, we get the following equation
\begin{equation}
{\bf{P}}_{n+1}^{(n+1)} = {\bf{\widehat P}}_{n+1}^{(n+1)},
{\bf{U}}_{n+1}^{(n+1)} = {\bf{\widehat U}}_{n+1}^{(n+1)}.
\end{equation}
Other procedures for this scenario is exactly the same as those in the previous proposed method. Finally, we can get the equivalent current ${\bf{J}}_{n+1}$ of the target surface $\gamma_{n+1}$ as
\begin{equation}
{{\bf{J}}_{n + 1}} = {{\bf{H}}_{n + 1}} - {\widehat {\bf{H}}_{n + 1}} = {{\bf{Y}}_{s_{n + 1}}}{{\bf{E}}_{n + 1}}.
\end{equation}

\section{Remarks Upon the Proposed Approach}
Let's assume objects embedded in multilayers with $m_1, m_2, ..., m_n$ segments on each interface and $N = m_1 + m_2 + ... + m_n$. Therefore, the PMCHWT formulation requires 2$N$ unknowns in the final system and all the geometrical fine details are directly included in the final system. However, for the proposed approach, the dimension of the final system is $m_n$, which only dependents on the unknowns residing on the outermost boundary $\gamma_n$ and is significantly smaller than that of the PMCHWT formulation. However, as shown by (\ref{Y2}), several matrix inversion and multiplying are required to construct the final system, which implies overhead for time cost. If many small scatters are involved as shown in [22], this overhead can be ignored compared with the overall time cost since that of the direct matrix inversion algorithm, like the LU decomposition, is small. Although overhead is required to construct the DSAO ${\bf{Y}}_{s_n}$, the computational performance improvement in terms of memory consumption and CPU time can still be significant as shown in the numerical examples in the next section. However, when large scatters are involved in the computational domain, direct construction of intermediate matrixes can be computationally  expensive. This problem can be mitigated using efficient direct solvers, like hierarchical matrix based direct algorithm \cite{BORM}, \cite{HJIAO}, Hierarchically Semi-Separable (HSS) solver \cite{HSS}, \cite{HSS2}. Acceleration of the proposed approach with those mentioned techniques is beyond the scope of this paper. We will report related results in the future.  Another merit of the proposed approach is that the condition of the final system can be improved compared with that of the PMCHWT formulation and can be similar to that of the EPA based techniques \cite{EPA}, \cite{DSAOARBSHAPEARRAY} since the geometry fine details are implicitly incorporated into the final system of the proposed approach.

\section{Numerical Results and Discussion}
All numerical examples in this paper are performed on a computer with Intel i7-7700 3.6 GHz CPU and 32G memory. All approaches except the Comsol are coded through the Matlab. All the codes of different approaches are running using one thread to make a fair comparison.
\subsection{A Single Dielectric Object}
\begin{figure}
	\begin{minipage}[h]{0.48\linewidth}\label{FIG6A}
		\centerline{\includegraphics[scale=0.5]{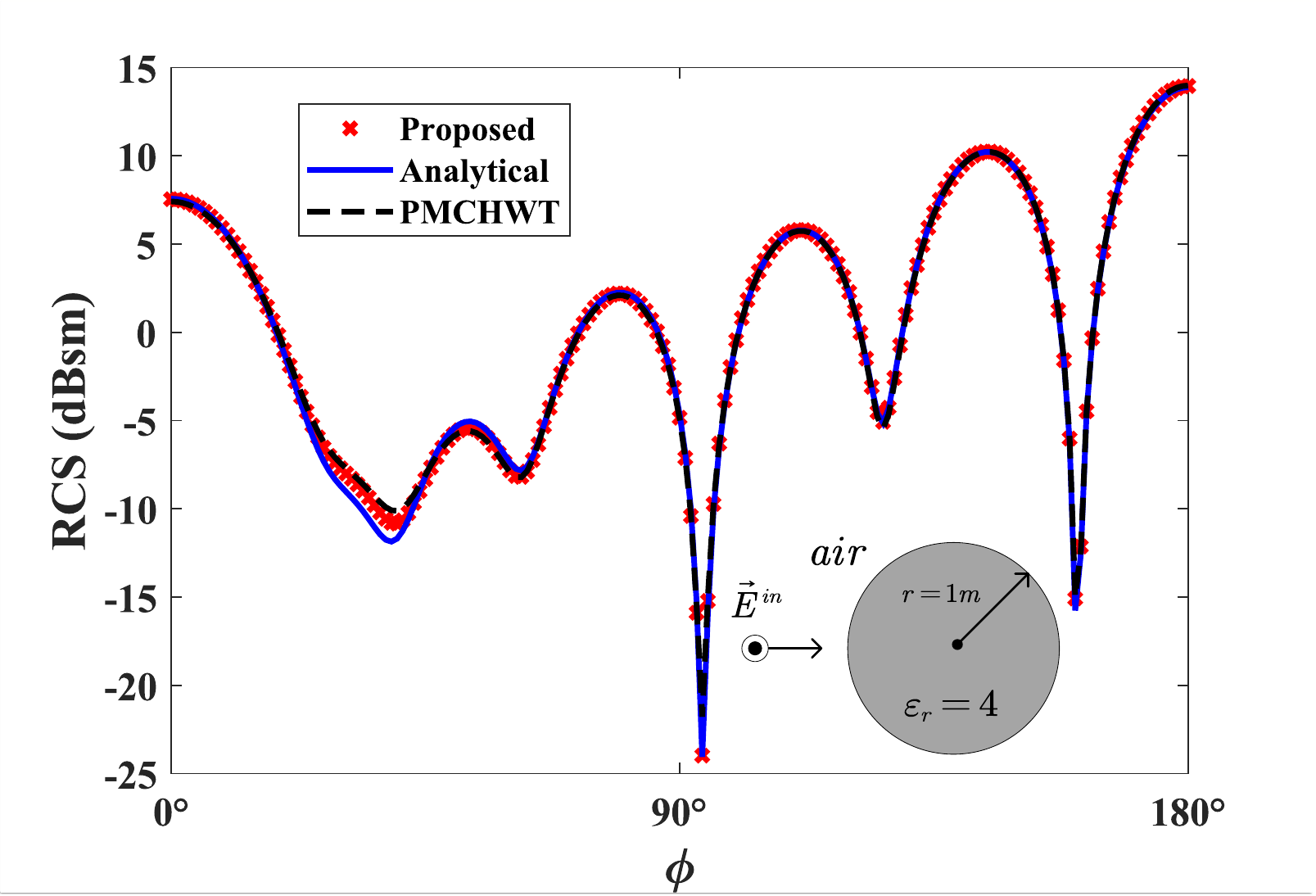}}
		\centerline{(a)}
	\end{minipage}
	\hfill
	\begin{minipage}[h]{0.48\linewidth}\label{FIG6B}
		\centerline{\includegraphics[scale=0.5]{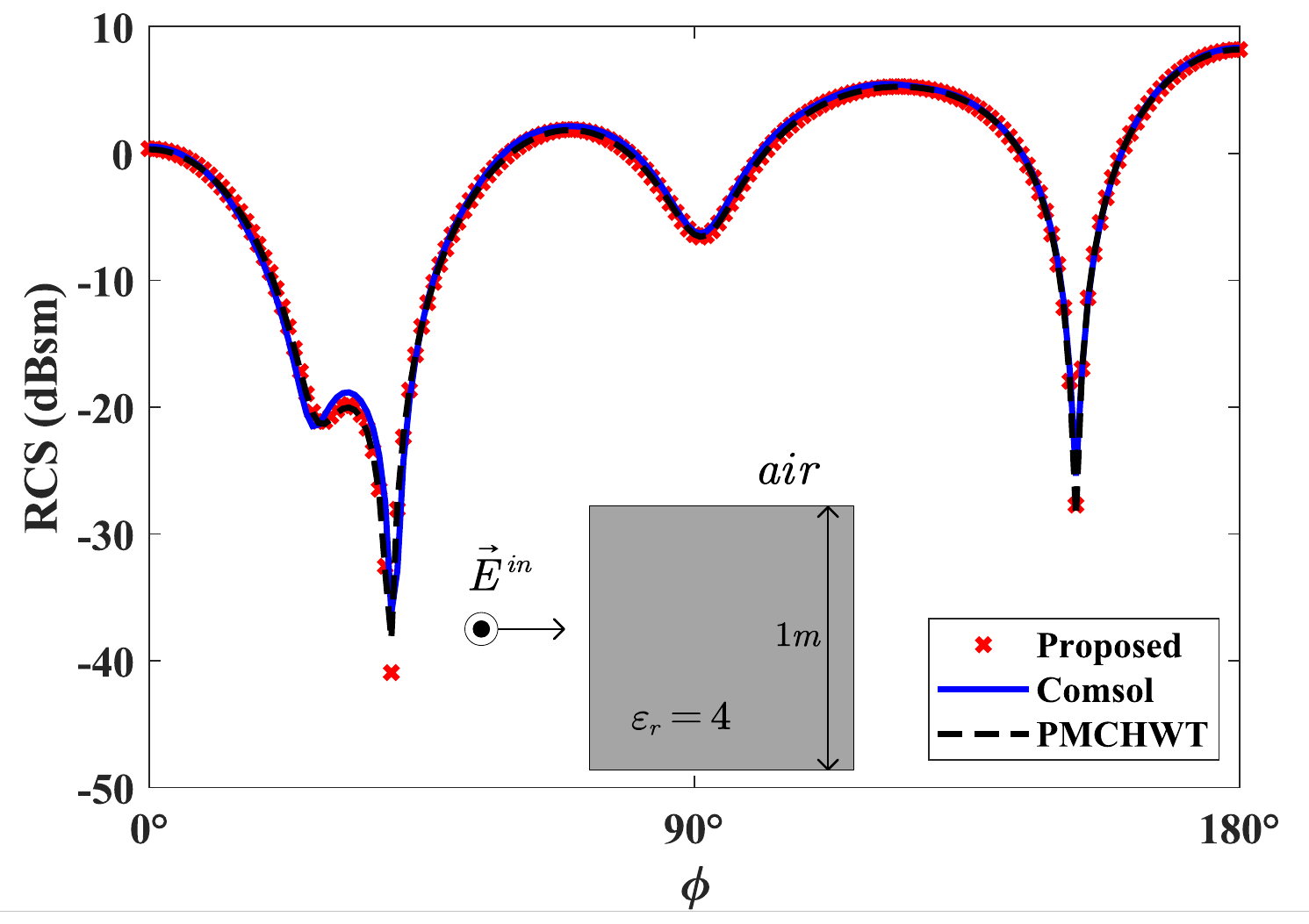}}
		\centerline{(b)}
	\end{minipage}
	
    \caption{(a) RCS obtained from the proposed method, the PMCHWT formulation and the analytical solution, and (b) RCS obtained from the proposed method, the PMCHWT formulation and the Comsol.}
    \label{RCS_single}
\end{figure}

We first consider two objects, an infinite long dielectric cylinder and cuboid with the relative permittivity $\varepsilon_{r} = 4$. The radius of the cylinder and the side length of the cuboid are both  $1\lambda_0$, where $\lambda_0$ is the wavelength in the free space. The incident plane wave is along the $x$-axis with $f=300$MHz and the averaged segment length is used as 0.05m to discretize the contour, which corresponds to $\lambda/10$, where $\lambda$ is the wavelength inside dielectric objects.

\begin{figure}
	\begin{minipage}[h]{0.48\linewidth}\label{FIG2C}
		\centerline{\includegraphics[scale=0.5]{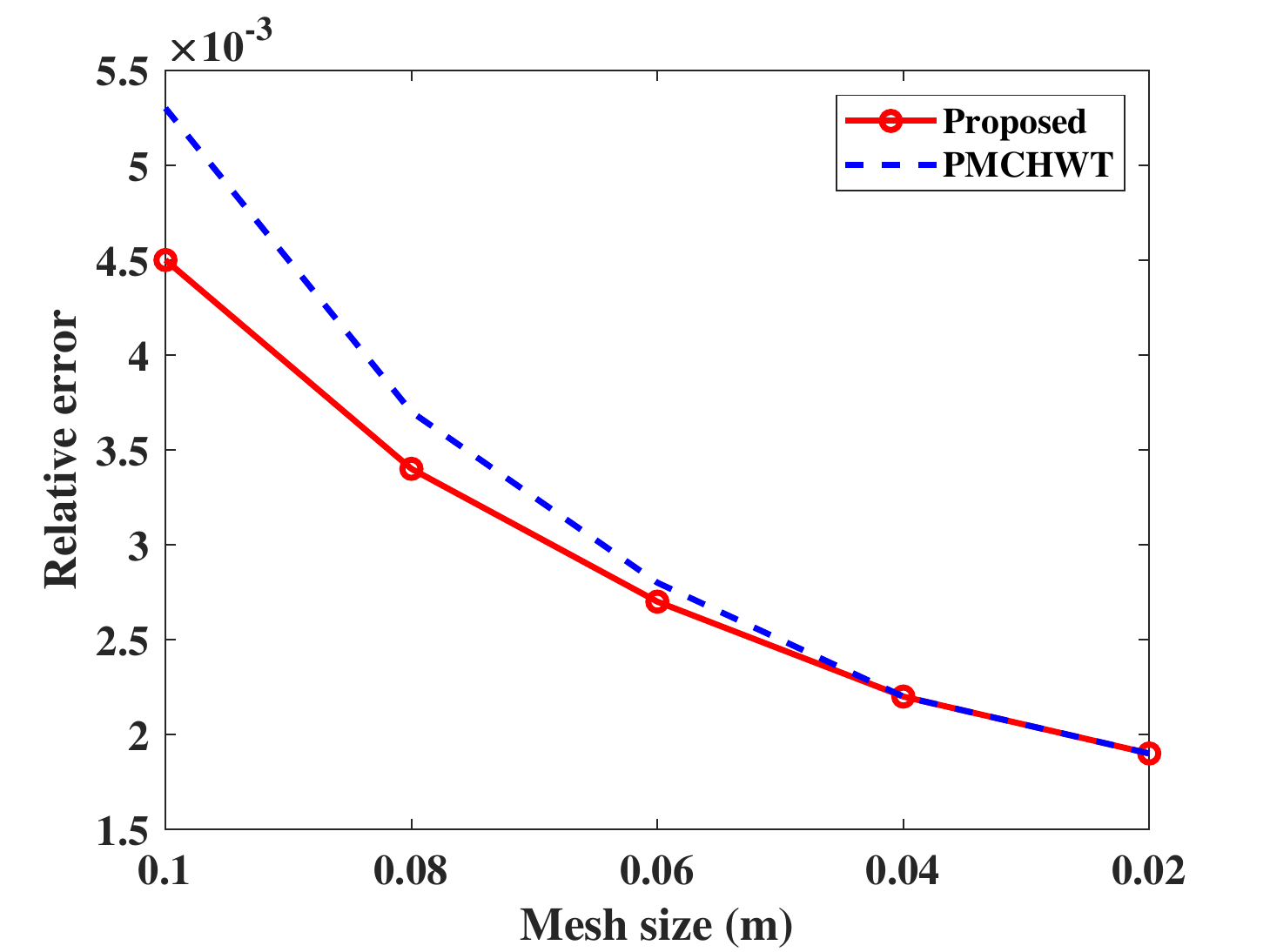}}
		\centerline{(a)}
	\end{minipage}
    \hfill
    \begin{minipage}[h]{0.48\linewidth}\label{FIG2C}
		\centerline{\includegraphics[scale=0.5]{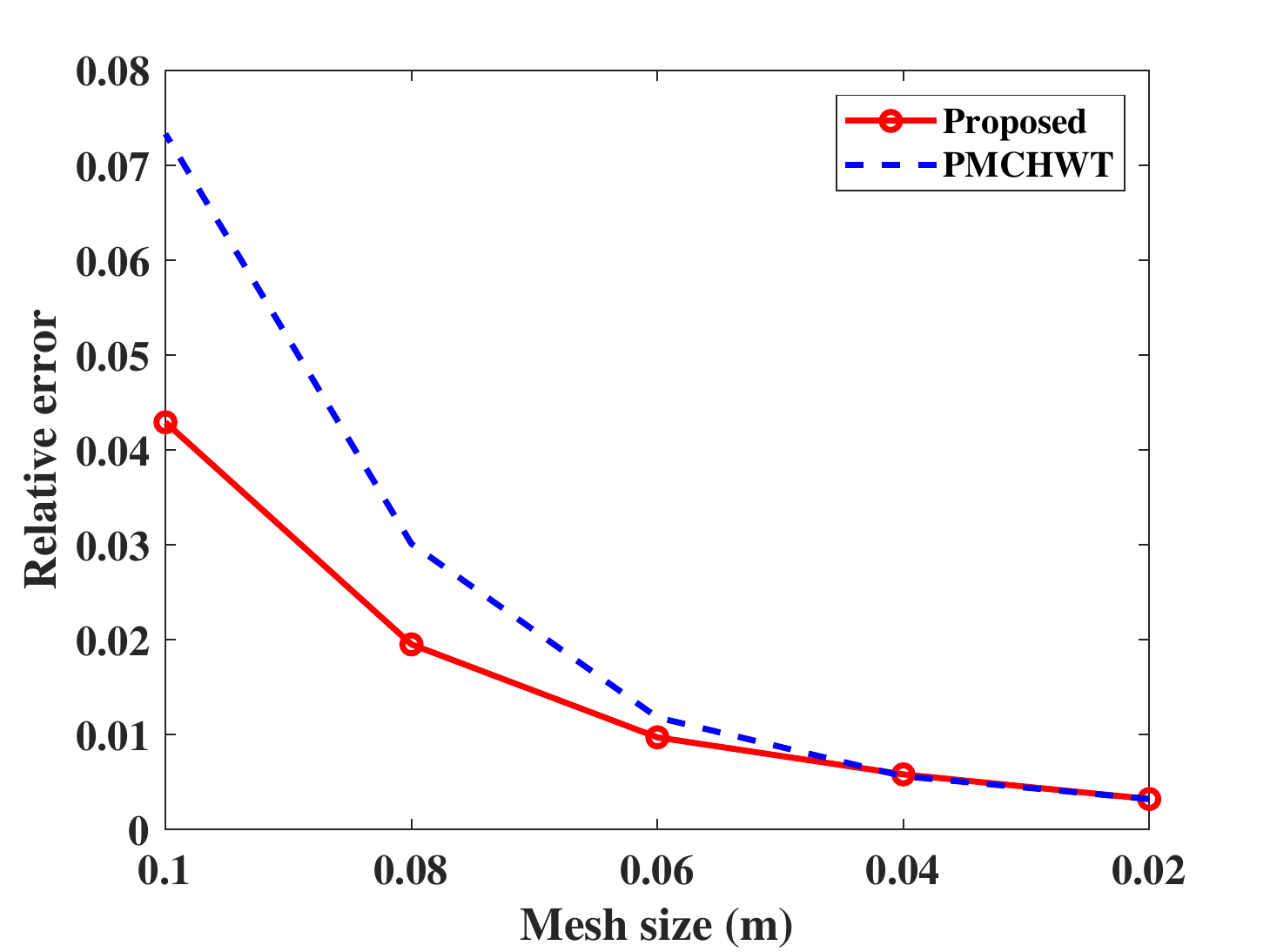}}
		\centerline{(b)}
	\end{minipage}

    \caption{Relative error of the proposed method and the PMCHWT formulation for (a) the dielectric cylinder and (b) the dielectric cuboid.}
    \label{RMS_single}
\end{figure}

Fig. \ref{RCS_single} shows RCS of the two dielectric objects obtained from the proposed method, the PMCHWT formulation and the analytical method or the Comsol. The reference RCS of the cuboid is obtained from the Comsol with extremely fine mesh. It is easy to find that the results obtained from the proposed approach for both objects agree well with those from the Comsol, the analytical method and the PMCHWT formulation. Therefore, the proposed approach can obtain accurate results for both smooth and non-smooth objects.

Then, we check the convergence property of the proposed method with mesh sizes. The relative error (RE) is  defined as
\begin{equation}
\text{RE}=\frac{\sum{_i \left \lVert \text{RCS}^{\text{cal}} \left( \phi _i \right) - \text{RCS}^{\text{ref}}\left( \phi _i \right) \right \rVert}^2}{\sum{_i \left \lVert \text{RCS}^{\text{ref}}\left( \phi _i \right) \right \rVert ^2}},
\end{equation}
where $ \text{RCS}^{\text{cal}} \left( \phi _i \right) $ denotes results calculated from the proposed approach or the PMCHWT formulation and $ \text{RCS}^{\text{ref}} \left( \phi _i \right) $ is the reference obtained from the analytical method or the Comsol. As shown in Fig. \ref{RMS_single}, it can be found that as mesh size decreases, the relative errors of both the PMCHWT formulation and the proposed approach decrease. When mesh size is relative large, say larger than $0.04$m, the relative error of the proposed approach is smaller than that of the PMCHWT formulation. Then, as we further decreases mesh size, both approaches reaches the same level of accuracy.

\begin{figure}
	\begin{minipage}[h]{0.48\linewidth}\label{FIG2A}
		\centerline{\includegraphics[scale=0.5]{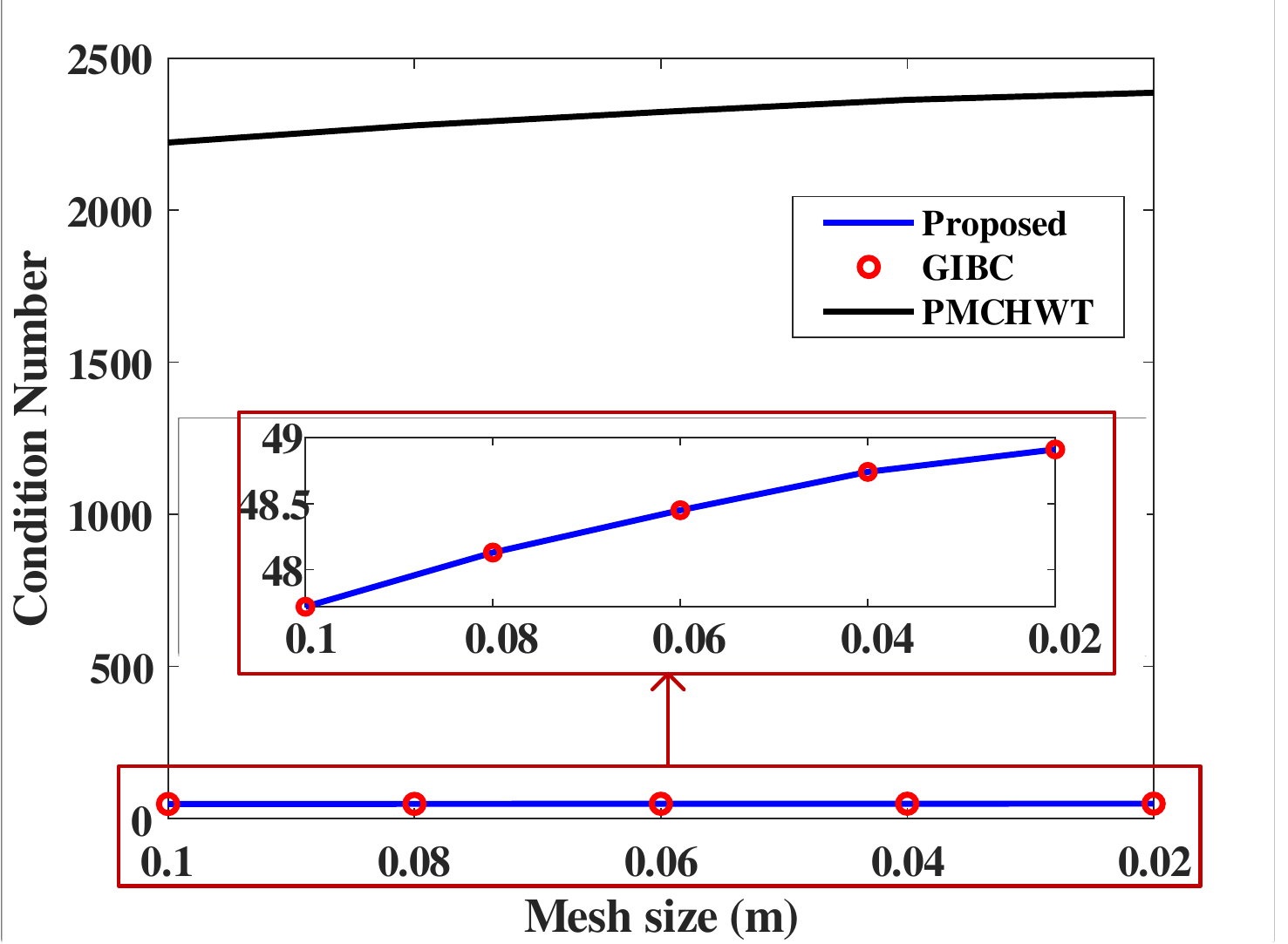}}
		\centerline{(a)}
	\end{minipage}
	\hfill
	\begin{minipage}[h]{0.48\linewidth}\label{FIG2B}
		\centerline{\includegraphics[scale=0.5]{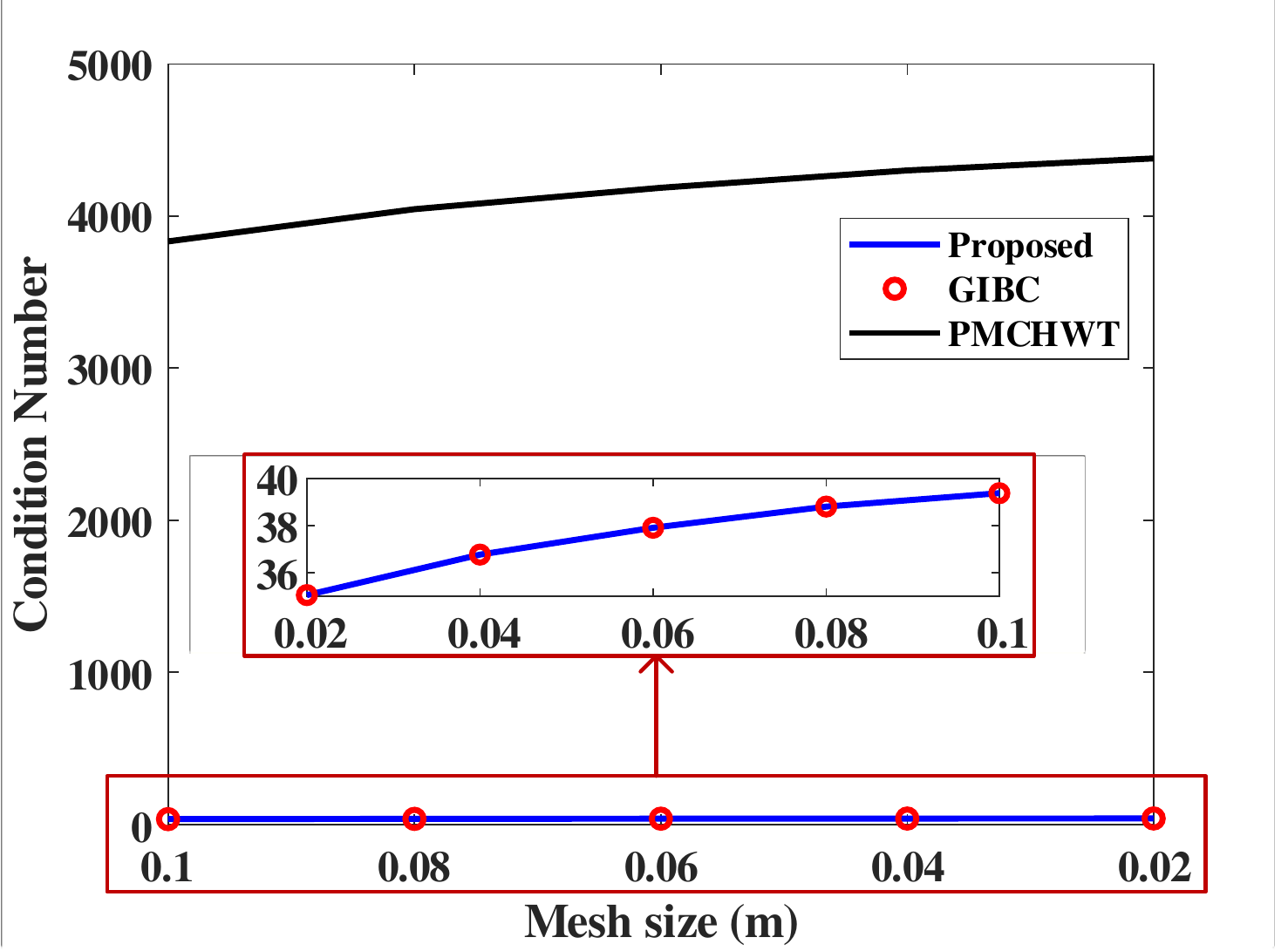}}
		\centerline{(b)}
	\end{minipage}
	\caption{The condition number of the final linear equations for the proposed approach, the GIBC formulation, and the PMCHWT formulation for (a) the cylinder and (b) the cuboid.}
	\label{CondNumber_single}
\end{figure}

In addition, we investigate the condition number of the final linear equations. For comparison purpose, the GIBC formulation is also included in our study.
As shown in Fig. \ref{CondNumber_single}, the condition number of the three approaches slightly increases as mesh size decreases. However, the condition number of the proposed formulation and the GIBC formulation is much smaller than that of the PMCHWT formulation. Take mesh size 0.1m for example. The condition number of the proposed formulation, the GIBC formulation, and the PMCHWT formulation is 47.72, 47.72, and 2222.41, respectively. It can be found that the proposed formulation is much better conditioning and then better convergence property than the PMCHWT formulation. It should be noted that only a single dielectric object is involved in each simulation. Therefore, only a half of count of unknowns for the proposed formulation compared with that of the PMCHWT formulation is required.

\subsection{A Single Layer Coated Cylindrical Conductor}
An infinitely long cylinder conductor coated with a layer dielectric medium is then considered. As shown in Fig. \ref{coppermodel+RCS}, the radius of the conductor and the coated medium are $10$ mm and $14$ mm. The conductivity of the conductor is 5.6 $\times 10^7$ $S/m$ and the relative permittivity ${\varepsilon}_r$ of the coated medium is $ 2.3$. The background medium is air. A plane wave incidents from the $x$-axis with the frequency of 30 GHz. The radar cross section (RCS) is compared with the proposed approach, the PMCHWT formulation, and the Comsol. The reference solution is obtained from the Comsol with extremely fine mesh.

\begin{figure}
	\begin{minipage}[h]{0.48\linewidth}\label{coppermodel+RCS}
		\centerline{\includegraphics[scale=0.5]{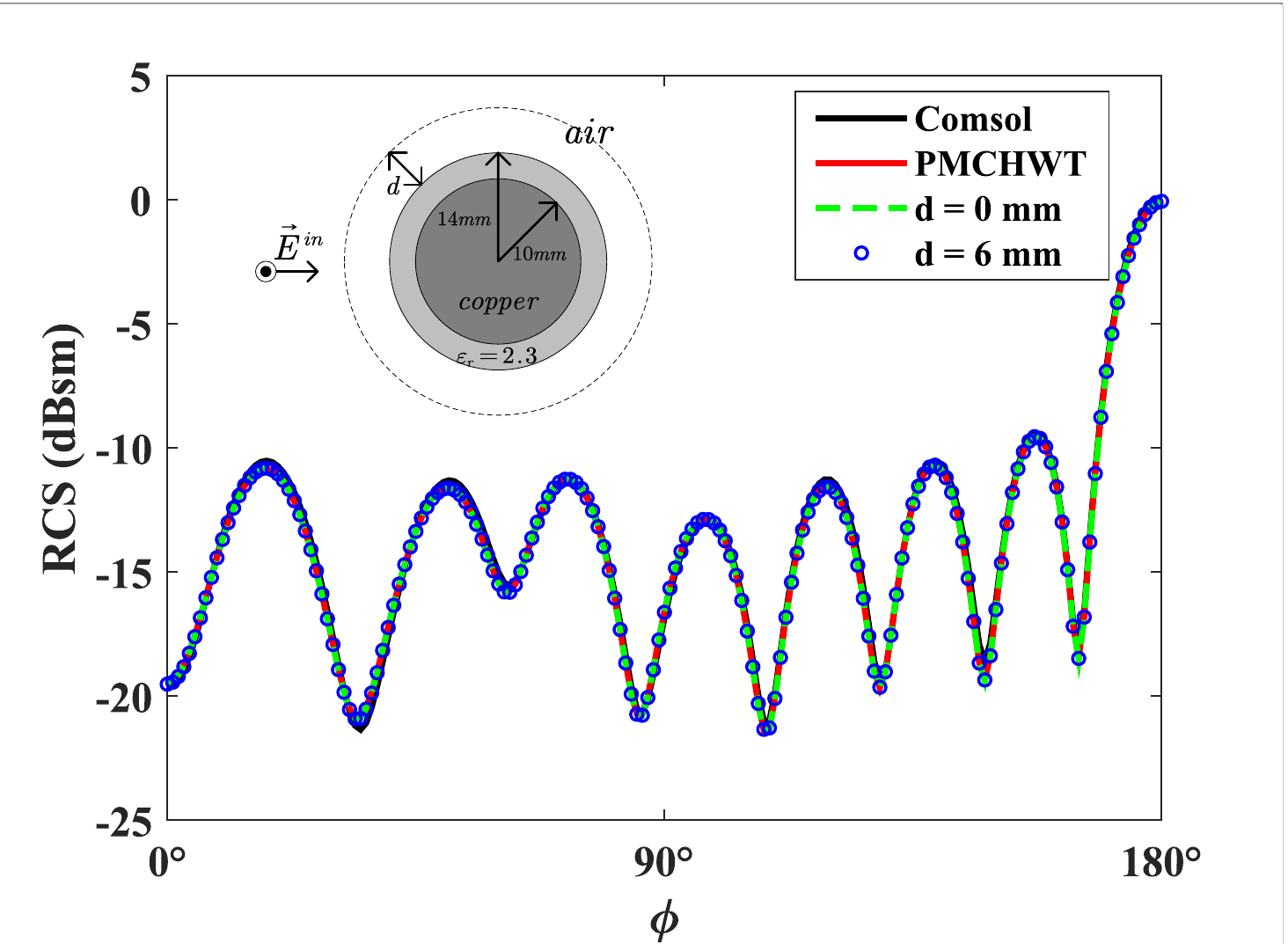}}
		\centerline{(a)}
	\end{minipage}
	\hfill
	\begin{minipage}[h]{0.48\linewidth}\label{copperRMS}
		\centerline{\includegraphics[scale=0.5]{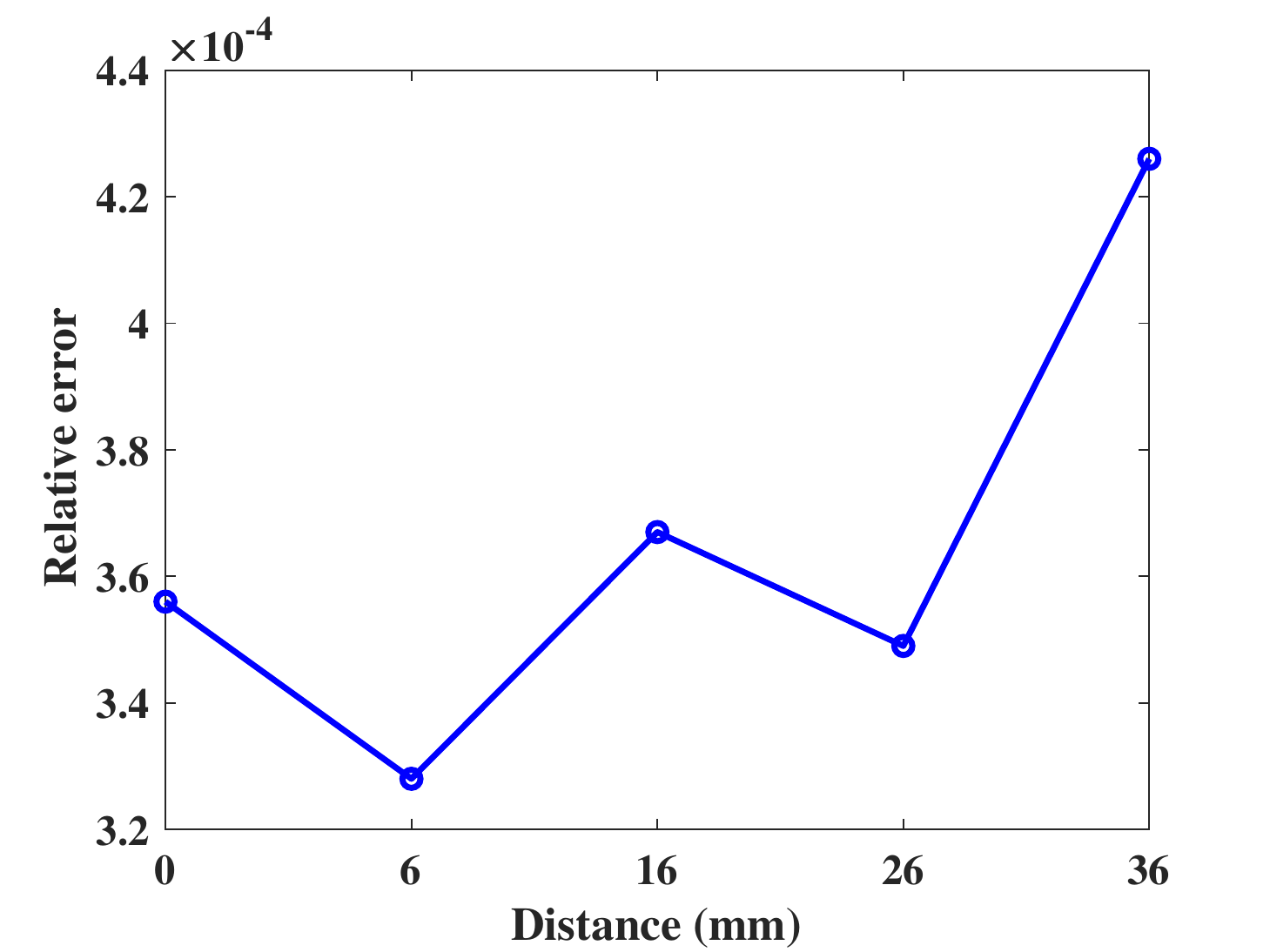}}
		\centerline{(b)}
	\end{minipage}
	\caption{(a) RCS obtained from the Comsol, the PMCHWT formulation, the proposed approach for a cylindrical conductor coated with a layer dielectric medium, and the proposed approach for the object with a fictitious boundary introduced in the exterior region, and (b) Relative error obtained from the proposed method with different $d$.}
	\label{coppermodel+RCS}
\end{figure}

\begin{table}[!t]
	\begin{center}
		\caption{COMPARISON OF THE CONDITION NUMBER BETWEEN THE PROPOSED APPROACH AND THE PMCHWT FORMULATION}
\begin{tabular}{c|c|c|c|c|c|c|c|c|c|c}
\toprule
\midrule
\multicolumn{11}{c}{Condition Number} \\
\hline
\multicolumn{2}{c|}{Proposed}
& \multirow{2}{*}{$\mathbf{P}_{1}^{\left( 1 \right)}$}
& \multirow{2}{*}{$\mathbf{\hat{P}}_{1}^{\left( 1 \right)}$}
& \multirow{2}{*}{$\mathbf{V}_{1}^{\left( 2 \right)}$}
& \multirow{2}{*}{$\mathbf{P}_{2}^{\left( 2 \right)}+\mathbf{F}_{2}^{\left( 2 \right)}\mathbf{P}_{1}^{\left( 2 \right)}$}
& \multirow{2}{*}{$\mathbf{\hat{P}}_{2}^{\left( 2 \right)}$}
& $\mathbf{L}_2+\mathbf{\hat{P}}_{2}^{\left( 2 \right)}\mathbf{Y}_{s_2} $
& \multirow{2}{*}{$\mathbf{P}_{3}^{\left( 3 \right)}+\mathbf{F}_{3}^{\left( 3 \right)}\mathbf{P}_{2}^{\left( 3 \right)}$}
& \multirow{2}{*}{$\mathbf{\hat{P}}_{3}^{\left( 3 \right)}$}
& {$\mathbf{L}_3+\mathbf{\hat{P}}_{3}^{\left( 3 \right)}\mathbf{Y}_{s_3}$} \\
\cline{1-2}
$d$ & $m_3$ &~&~&~&~&~&$\left( =\mathbf{V}_{2}^{\left( 3 \right)} \right)$&~&~& (final matrix) \\
\hline
0 &176 &1 &25.09 &25.08 &26.51 &64.43 &131.36 & - & - & - \\
\hline
6 &252 &1 &25.09 &25.08 &26.51 &64.43 &131.36 &50.50 &41.32 &69.26 \\
\hline
16 &380 &1 &25.09 &25.08 &26.51 &64.43 &131.36 &87.35 &47.11 &11.44 \\
\hline
26 &504 &1 &25.09 &25.08 &26.51 &64.43 &131.36 &51.19 &51.19 &65.07 \\
\hline
36 &628 &1 &25.09 &25.08 &26.51 &64.43 &131.36 &67.86 &330.67 &497.02 \\
\hline
\multicolumn{2}{c|}{PMCHWT}& \multicolumn{9}{c}{7436.56} \\
\hline
\bottomrule
\end{tabular}
\end{center}
\end{table}

Two scenarios are considered for the proposed approach: the object with or without an extended fictitious boundary in the background medium as shown in Fig. \ref{coppermodel+RCS}. $d$ is the distance between the outermost boundary and the fictitious boundary. The RCS with $d$ = 0 mm, which corresponds to the case that results obtained without the fictitious boundary, $d$ = 6 mm, the PMCHWT formulation and the Comsol are shown in Fig. \ref{coppermodel+RCS}(a).
This fictitious boundary can be selected arbitrarily. There is only one implication that the boundary should be enclosed and include the whole object.
Results obtained from the proposed approach with $d$ = 0 mm, 6 mm agree well with the reference solution. It demonstrates that the proposed approach can accurately model the multilayer embedded objects.

Then, to study the numerical stability of the proposed method with respect to $d$, we calculated the relative errors with different $d$. As shown in Fig. \ref{coppermodel+RCS}(b), the relative errors for all cases are quite small, which implies that the numerical stability of the proposed approach is quite good. Table 1 shows the number of unknowns and the condition number required by our proposed approach and the PMCHWT formulation with different $d$. The table lists all the condition numbers of the intermediate matrixes requiring inversion in our proposed approach. As shown in Table I, all the condition numbers of intermediate matrixes are quite small, which means that quite low cost is required to calculate their matrix inversion when iterative algorithms are used. As $d$ increase, the condition number of the final system seems to grow larger. Similar to the previous numerical example, the final condition number of the linear equation of the proposed approach is much smaller than that of the PMCHWT formulation.  It should be noted that as $d$ increases, the number of unknowns on the fictitious boundary increases as shown in first column of Table 1. Therefore, a balance with respect to the count of unknowns should be made for $d$.

\subsection{A Complex Structure}
\begin{figure}
	\centerline{\includegraphics[scale=0.6]{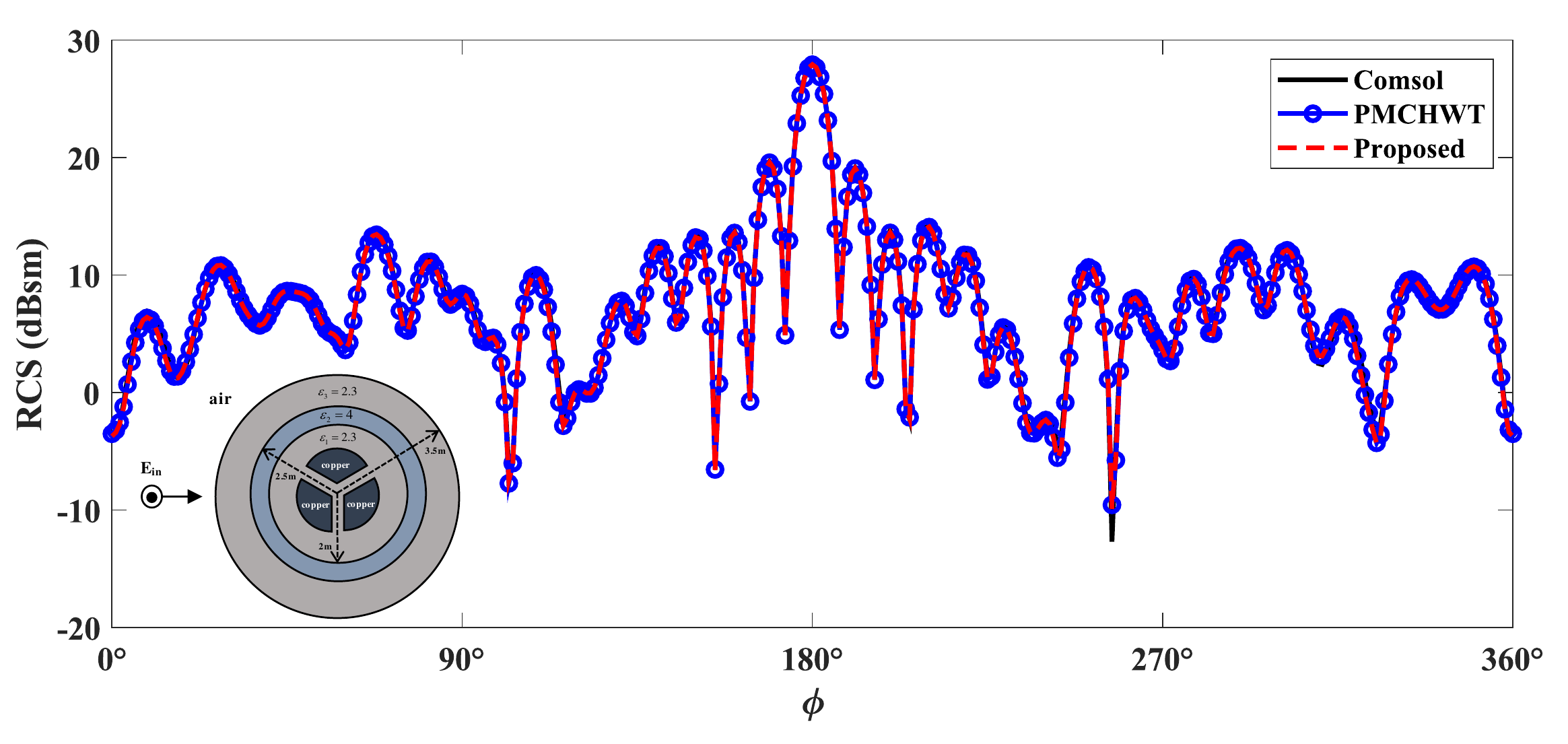}}
	\caption{Configurations of the multilayer embedded objects and RCS from Comsol, PMCHWT and the proposed method.}
	\label{ComplexRCS+Model}
\end{figure}

\begin{table}[!t]
	\begin{center}
		\caption{COMPARISON OF THE OVERALL COUNT OF UNKNOWNS, THE COUNT OF FLIP-FLOP OPERATIONS AND TIME COST BETWEEN THE PROPOSED APPROACH AND THE PMCHWT FORMULATION}
		\begin{tabular}{c c c c}
			\toprule
			\midrule
			\multirow{1}{*}{Metric}    & PMCHWT & Proposed  & Ratio\\
            \cmidrule(r){1-2}\cmidrule(l){2-4}
			\multicolumn{1}{l}{Count of unknowns}     & 4188  & 736  & 0.18 \\
			\cmidrule(r){1-2}\cmidrule(l){2-4}
			\multicolumn{1}{l}{Count of flip-flop operations}     & $7.347\times 10^{10}$  & $5.107\times 10^{9}$ & 0.07   \\
			\cmidrule(r){1-2}\cmidrule(l){2-4}
			\multicolumn{1}{l}{Filling matrix [s]}     & 43.52  & 36.62  & 0.84 \\
			\midrule
            \multicolumn{1}{l}{solving matrix [s]}     & 1.6  & 0.43  & 0.27 \\
			\midrule
            \multicolumn{1}{l}{Overall time cost [s]}     & 45.12  & 37.05  & 0.82 \\
			\midrule
			\bottomrule
		\end{tabular}
	\end{center}
\end{table}

\begin{table}[!t]
	\begin{center}
		\caption{COMPARISON OF THE CONDITION NUMBER BETWEEN THE PROPOSED APPROACH AND THE PMCHWT FORMULATION}
\begin{tabular}{l|c|c|c|c|c|c|c}
\toprule
\midrule
\multicolumn{8}{c}{Condition Number} \\
\hline
\multicolumn{7}{c|}{Proposed} & PMCHWT\\
\hline
\multirow{2}{*}{The 1st Layer} & $\mathbf{P}_{1_1}^{\left( 1 \right)}$ &1.02  & $\mathbf{P}_{1_2}^{\left( 1 \right)}$ & 1.02 & $\mathbf{P}_{1_3}^{\left( 1 \right)}$ & 1.02 &\multirow{6}{*}{24745.34} \\
\cline{2-7}
& $\mathbf{\hat{P}}_{1_1}^{\left( 1 \right)}$ & 41.03 & $\mathbf{\hat{P}}_{1_2}^{\left( 1 \right)}$ & 40.81 & $\mathbf{\hat{P}}_{1_3}^{\left( 1 \right)}$ & 41.03 \\
\cline{1-7}
The 2nd Layer & $\mathbf{V}_{1}^{\left( 2 \right)}$ & 75.75 & $\mathbf{P}_{2}^{\left( 2 \right)}+\mathbf{F}_{2}^{\left( 2 \right)}\mathbf{P}_{1}^{\left( 2 \right)}$ & 51.40& $\mathbf{\hat{P}}_{2}^{\left( 2 \right)}$ & 42.45 \\
\cline{1-7}
The 3rd Layer & $\mathbf{V}_{2}^{\left( 3 \right)}$ & 32.65 & $\mathbf{P}_{3}^{\left( 3 \right)}+\mathbf{F}_{3}^{\left( 3 \right)}\mathbf{P}_{2}^{\left( 3 \right)}$ & 62.67& $\mathbf{\hat{P}}_{3}^{\left( 3 \right)}$ & 55.48 \\
\cline{1-7}
The 4th Layer & $\mathbf{V}_{3}^{\left( 4 \right)}$ & 72.61 & $\mathbf{P}_{4}^{\left( 4 \right)}+\mathbf{F}_{4}^{\left( 4 \right)}\mathbf{P}_{3}^{\left( 4 \right)}$ & 87.89& $\mathbf{\hat{P}}_{4}^{\left( 4 \right)}$ & 82.57 \\
\cline{1-7}
{Solving $\mathbf{E}_4$} & \multicolumn{3}{c|}{$\mathbf{L}_4+\frac{1}{2}\mathbf{\hat{P}}_{4}^{\left( 4 \right)}\mathbf{Y}_{s_4}$ (final matrix)} & \multicolumn{3}{c|}{198.65}  \\
\hline
\bottomrule
\end{tabular}
\end{center}
\end{table}

We finally consider an object with a complex geometry as shown in Fig. \ref{ComplexRCS+Model}. The innermost objects are three sector-shaped copper conductors with an angle of 120 degrees and a radius of 1 m. The relative permittivity of the second, third and fourth layer dielectric cylindrical media are 2.3, 4.0, 2.3, respectively.  This object contains both a multilayer structure and multiple scatters inside the layered media.

The RCS is calculated from the proposed approach, the PMCHWT formulation, the Comsol as shown in the Fig. \ref{ComplexRCS+Model}. Results from the three approaches show excellent agreement with each other. It demonstrates that the proposed approach can indeed accurately model the complex objects embedded in multilayers.

Table 2 shows the overall count of unknowns, the count of flip-flop operations, time cost of the proposed approach and the PMCHWT formulation. The ratio in the last column of Table 2 is defined as the ratio of the quantity of the proposed approach to that of the PMCHWT formulation. As shown in the second row of Table 2, the proposed approach only needs 736 unknowns to solve this complex problem compared with 4188 unknowns for the PMCHWT formulation. Its ratio is 0.18, which means only quite a small fraction of amount of unknowns is required in the proposed approach. This is becasue that the PMCHWT formulation requires both the electric and magnetic current densities on all the interfaces of different homogenous media. However, in the proposed formulation only single electric current density on the outermost boundary of objects is required. Therefore, the overall count of unknowns can be significantly reduced. Moreover, we count the overall flip-flop operations for the two approaches as shown in the third row of Table 2. The count of flip-flop operations of our proposed approach is only $7\%$ of that of the PMCHWT formulation. The time cost for both approaches are shown from the fourth to sixth rows of Table 2. The time cost for the PMCHWT formulation requires 45.12 seconds to solve this problem. However, only 37.05 seconds are required for the proposed approach. Its ratio is 0.82. The improvement for the time cost and the count of unknowns is not a linear scale. This is because that overhead is required to construct the DSAO $\mathbf{Y}_n$. However, as shown in Table 2, the time cost of matrix filling and solving linear equations is reduced compared with that of the PMCHWT formulation. As the number of interfaces increases, the performance improvement in terms of time cost and count of unknowns would be more significant.

As shown in Table 3, the condition numbers required by our proposed approach for all intermediate matrixes when we construct $\mathbf{Y}_n$. Similar observations with the previous numerical example can be obtained in this example. The condition number of the final linear equation in the PMCHWT formulation is 24,745. However, the condition number for the proposed formulation is only 198. Therefore, the proposed approach shows much better conditioning and then convergence property compared with the PMCHWT formulation.

\section{Conclusion}
In this paper, we proposed a novel and unified single-source SIE to model multilayer embedded objects. The proposed approach only needs a single electric current density on the outermost boundary of objects, which can be derived by recursively applying the equivalent theorem on each boundary from inner to exterior regions. The surface electric field is related to the magnetic field on the outermost boundary through a DSAO. Then, with combining the EFIE with the equivalent current density, we can accurately solve the scattering problems by objects embedded in multilayers. Numerical results demonstrate that the proposed approach can significantly improve the performance in terms of the count of unknowns and time costs, which implies great potential for electromagnetic analysis of complex multilayer embedded objects. Currently, development of the proposed approach into three dimensional general scenarios is in progress. We will report more results on this topic.

\section*{Acknowledgement}
This work was supported in part by the National Natural Science Foundation of China through Grant 61801010, in part by Beijing Natural Science Foundation through Grant 4194082.



\end{document}